\documentclass[prb,superscriptaddress,twocolumn,floatfix,notitlepage,citeautoscript]{revtex4-2}
\usepackage[T2A]{fontenc}
\usepackage{mmap}
\usepackage{amssymb}
\usepackage{amsmath}
\usepackage{color}
\usepackage[usenames,dvipsnames]{xcolor}
\usepackage{hyperref}
\usepackage{textcomp}
\hypersetup{colorlinks=true,linkcolor=blue,urlcolor=blue,citecolor=blue}
\usepackage{natbib}
\usepackage{graphicx}
\usepackage{dsfont}
\usepackage{mathrsfs}

\begin{document}
\title{Controlled Polarization Switch in a Polariton Josephson Junction}

\author{Valeria A. Maslova}
\affiliation{National Research Nuclear University MEPhI (Moscow Engineering Physics Institute), 115409 Moscow, Russia}
\affiliation{C.H.O.S.E. (Center for Hybrid and Organic Solar Energy), Electronic Engineering Department, University of Rome Tor Vergata, Via del Politecnico 1, 00118, Rome, Italy}
\affiliation{Emerging Technologies Research Center, XPANCEO, Internet City, Emmay Tower, Dubai, UAE}

\author{Nina~S.~Voronova}
\email{nsvoronova@mephi.ru}
\affiliation{National Research Nuclear University MEPhI (Moscow Engineering Physics Institute), 115409 Moscow, Russia}
\affiliation{Russian Quantum Center, Skolkovo IC, Bolshoy boulevard 30 bld. 1, 121205 Moscow, Russia}

\begin{abstract}
The interaction between a particle's spin and momentum---known as spin-orbit (SO) coupling---is the cornerstone of modern spintronics.
In Bose-Einsten condensates of ultracold atoms, SO coupling can be implemented and precisely controlled experimentally; photonic systems, on the other hand, possess an {\it intrinsic} SO interaction due to the longitudinal-transverse splitting of the photon modes.
In this work, we focus on such spinor, SO-coupled exciton-polariton condensates on a ring, where the strength of the synthetic magnetic field is controlled by the geometrical dimensions of the structure.
Inspired by recent experiments, we investigate the dynamics of a weakly-nonlinear four-mode bosonic Josephson junction within this geometry. We discover a narrow parameter range in which the interplay of the tunneling dynamics with polariton-specific SO coupling leads to a new regime, with dynamical switching of the fluid's circular polarization degree to the opposite, along the entire ring or on just one of its halves.
Our results demonstrate polariton condensates in ring configurations as excellent candidates for all-optical controllable spin-switch applications, with prospects for scalability and observing non-trivial polarization patterns.
\end{abstract}

\maketitle

\section{Introduction}
In modern technology, spin-orbit (SO) effects in semiconductors and other materials are being actively explored for spintronics applications, which aim to utilize the electron's spin in addition to its charge for information processing~\cite{RMP_spintronics}.
Different from electrons, ultracold atoms don't have intrinsic SO interaction that couples their spin to center-of-mass motion. Instead, it is engineered using lasers both for neutral bosonic~\cite{Galitsky,Lin_SO_BECS} and fermionic gases~\cite{Zhang,Zwierlein}, typically with counter-propagating Raman beams that couple two internal atomic states (acting as a pseudospin) to the atom's momentum.
In this case, the synthetic magnetic field that couples the spin of a particle to its orbital motion is of the Rashba $\hat{\mathcal{H}}_{\rm R} \propto k_x \hat\sigma_y - k_y\hat\sigma_x$ and Dresselhaus $\hat{\mathcal{H}}_{\rm D} \propto k_x \hat\sigma_x - k_y\hat\sigma_y$ types, both linear in the particle's momentum. This brings about a variety of condensed-matter phenomena, such as topological phases and quantum Hall effect~\cite{Zhang_RMP}, alters the single-particle dispersion~\cite{Lin_SO_disp,Galitsky}, and leads to the so-called spin-Hall effect~\cite{Hirsch}. The interplay between spin dynamics and particles' motion leads to novel collective phases in spinor Bose-Einstein condensates (BECs)~\cite{Zhang_prl}. In a bosonic Josephson junction (BJJ)~\cite{Smerzi_prl, albiez2005direct, Levy, Oberthaler_prl}, this coupling intertwines the tunneling dynamics with the internal spin oscillations, giving rise to spin Josephson effect~\cite{Zhang_bjj, citro_epj}.

Another type of systems possessing intrinsic SO interaction is based on photonic platforms. In particular, liquid-crystal microcavities provide means to implement the RD type of coupling and realize the dispersion engineering similar to SO-coupled atomic BECs~\cite{Szczytko2019, Szczytko2022, gao_natcomm2022, sedov_prr2024}. On the other hand, to enable photons to interact and study their collective behavior, one needs to consider exciton-polaritons. Those are hybrid quasiparticles formed from the strong coupling of an exciton (a bound electron-hole pair in a semiconductor) and a cavity photon~\cite{microcavities}. Polaritons are bosons that can form condensates at much higher temperatures than ultracold atoms~\cite{kasprzak2006bose,byrnes2014exciton}. They possess both the pseudospin (inherited from the exciton spin and photon polarization) and a natural, built-in SO interaction arising from the transverse-electric (TE) -- transverse-magnetic (TM) splitting of the cavity modes, different from the Rashba and Dresselhaus couplings:
\begin{equation}\label{TE-TM}
    \hat{\mathcal{H}}_{\rm TE-TM} = \begin{pmatrix}
        0 & \beta (k_x-ik_y)^2\\
        \beta (k_x+ik_y)^2 & 0
    \end{pmatrix},
\end{equation}
where $\beta = (\hbar^2/4) (m_{\rm TM}^{-1}- m_{\rm TE}^{-1})$ and $m_{\rm TE(TM)}$ are the polariton effective masses associated with the two dispersions.
The pseudospin structure of polaritons gives rise to a plethora of effects that deserve special attention~\cite{shelykh2010review}. Direct consequences of the TE-TM splitting include the optical spin-Hall effect~\cite{OSHE,leyder}, anomalous Hall drift~\cite{Gianfrate_nature}, self-induced Larmor precession~\cite{Gnusov_Larmor}, and various topological phenomena~\cite{Amo_omex}.
Surprisingly, however, despite extensive studies on SO-coupled atomic BJJs~\cite{Zhang_bjj, citro_epj,polls_pra2014,fu_pra92,liu_2023},  a similar exploration of their polariton counterparts with TE-TM SO coupling is notably absent. The only prior study on polariton BJJs that considered pdeudospins~\cite{shelykh2008josephson} used a phenomenological spin-flip constant and neglected the interplay of external and internal Josephson effects.

\begin{figure}[b!]
\centering\includegraphics[width=\linewidth]{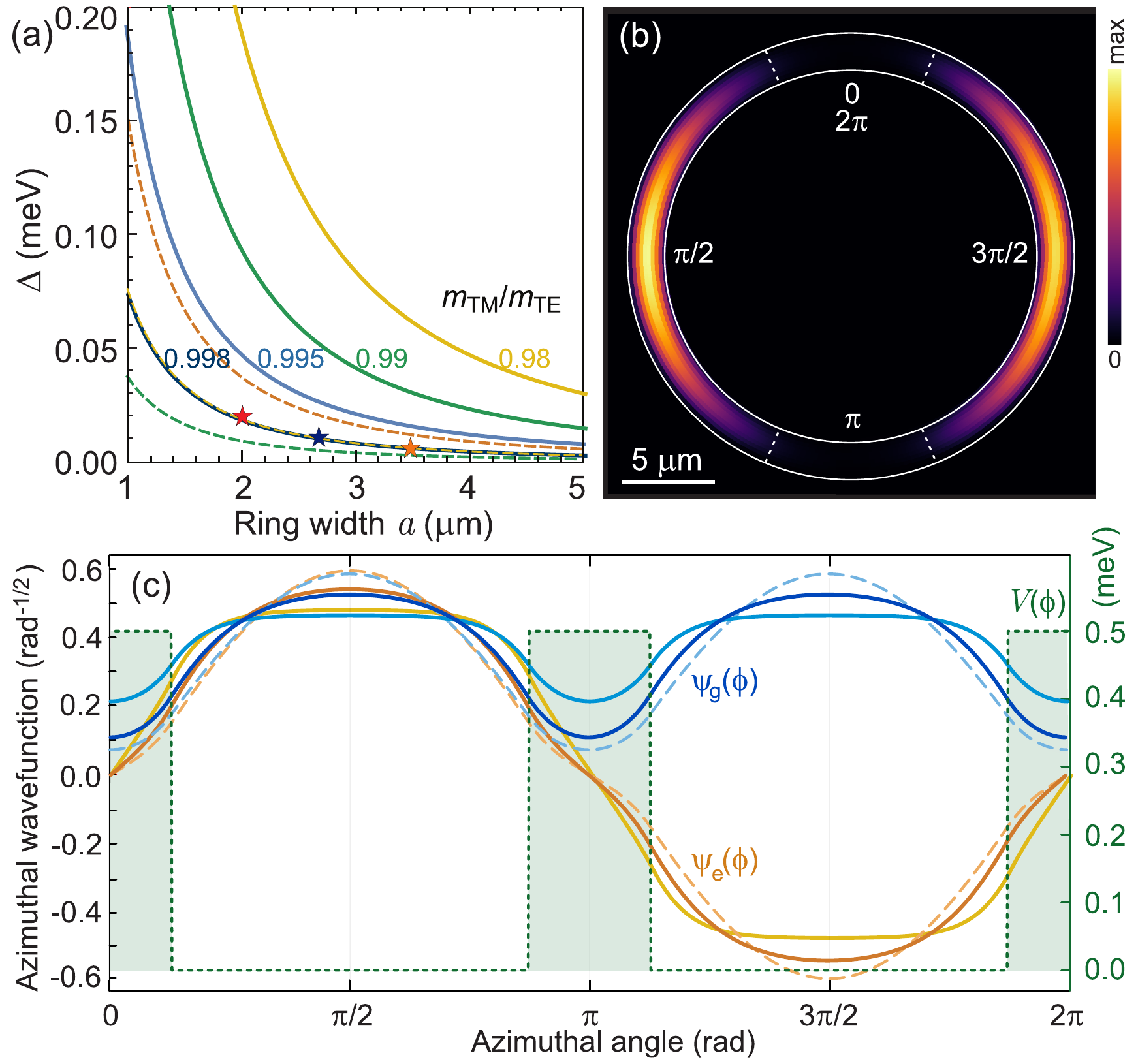}
\caption{\textbf{Geometry of the system.}
(a) Solid lines: TE-TM splitting vs. the ring width for different ratios $m_{\rm TM}/m_{\rm TE}$ as marked, for $m = 10^{-5}m_0$ ($m_0$ is the free electron mass). Dashed lines: same for $m = 10^{-4}m_0$ and $m_{\rm TM}/m_{\rm TE}=0.99$ (green), 0.98 (yellow) and 0.96 (orange). Stars indicate the values used in this work.
(b) Sketch of the polariton ring, $r_1=10~\mu$m, $r_2=12~\mu$m, with the two Josephson junctions dividing it effectively in two potential wells.
Color (in arb. units) shows the intensity map of the symmetric initial state (no particle imbalance).
(c) Right axis, dotted line: the azimuthal potential along the ring circumference shown in (b), of a height $V = 0.5$~meV and width $\alpha = 0.8$~rad. Left axis: the two lowest eigenstates $\psi_g$ (blue) and $\psi_e$ (yellow) of the GPE, for $g = 1~\mu$eV~$\mu$m$^2$ and the average one-component density $\rho = 50~\mu$m$^{-2}$ (dark solid lines) and $200~\mu$m$^{-2}$ (light solid lines). Thin dashed lines show the linear limit $g=0$.
}
\label{fig1}
\end{figure}

The SO coupling strengths in polariton systems are largely determined by the specific material and details of its growth~\cite{panzarini}, and are only slightly adjustable in the laboratory. Recently, however, ring-shaped polariton BECs have been experimentally implemented~\cite{lukoshkin2018persistent,mukherjee2019oscillation,mukherjee2021dynamics}, including rings with weak links that behave as Josephson junctions~\cite{barrat2024stochastic,voronova_JJ}.
In ring geometry, the TE-TM splitting is enhanced due to confinement in the radial direction~\cite{meijer} [see Fig.~\ref{fig1}(a)], leading to emergence of topological effects \cite{kozin2018topological,zezyulin2018chiral}, the pseudomagnetic field effect \cite{mukherjee2021dynamics}, and spontaneous symmetry breaking~\cite{chestnov2023symmetry}. Here, we focus on the dynamics of exciton-polariton condensate in ring geometry with two weak links, a system that realizes the SO-coupled photonic BJJ. The goal of this work is to explore the physics analogous to the spin Josephson effect in atomic BECs, which are produced by the interplay between extrinsic tunneling and TE-TM SO coupling.
Our analysis reveals qualitatively new phenomena, different from previously reported BJJ dynamics. Around a critical value of the condensate degree of circular polarization (DCP), the tunneling oscillatory dynamics of the double-well BJJ is suppressed. At the same time, under specific conditions, the polarization in one half-ring switches to the opposite, while in the other half the DCP can switch, oscillate or stay self-localized around the initial value. We present a diagram of dynamical regimes with experimentally adjustable parameters.

\section{Model}
The system under study is a thin ring  (with the inner and outer radii $r_1$ and $r_2$, respectively) divided in two halves by the two Josephson junctions, as shown in Fig.~\ref{fig1}(b). The potential barriers creating the junctions represent the so-called weak links: regions of space of the size of order of the condensate healing length $\xi=\hbar/\sqrt{2mg\rho}$, where the condensate density is largely suppressed. In the above, $m^{-1} = (m_{\rm TE}^{-1} + m_{\rm TM}^{-1})/2$ is the average polariton effective mass, $g$ the interaction constant, and $\rho$ the average one-component density. Importantly, we choose to work in the weakly interacting limit, so that $\xi\gg a$ ($a = r_2-r_1$ is the ring width), while it is still smaller than the ring circumference, ensuring the  weak links creation.
The three-dimensional potential of the problem can be represented as the sum of a box potential in the radial direction $r_1<r<r_2$ and the azimuthal potential $V(\phi)$ shown in Fig.~\ref{fig1}(c), with the two rectangular barriers of the height $V$ and the angular width $\alpha$. The tight radial confinement $a \ll (r_2+r_1)/2$ allows to truncate the problem to one dimension (1D), describing the azimuthal behavior of the order parameter (for details, see Appendix~\ref{ssec1}).

Since the two barriers effectively divide the ring into a doubly-connected two-well condensate, in the scalar case the dynamics of the order parameter is that of the BJJ (this case is discussed in Appendix~\ref{ssec2}; for a review, see~\cite{gati2007bosonic}).
In our work we are however interested in the interplay of the BJJ behavior and the pseudospin dynamics in presence of SO interaction for the polariton spinor $\Psi = (\Psi_\uparrow,\Psi_\downarrow)^T$. The Hamiltonian describing azimuthal motion on a thin ring in presence of the TE-TM splitting reads
\begin{equation}\label{hamiltonian_spinor}
    \hat{\mathcal{H}} = \! \begin{pmatrix}
    \! - \frac{\hbar^2 \partial^2_{\phi\phi}}{2mR^2}  \!+\! V(\phi) \!+\! \tilde{g} |\Psi_\uparrow|^2 & \Delta e^{-2i\phi} \\[5pt]
    \Delta e^{2i\phi} & \!\!\!\!  - \frac{\hbar^2\partial^2_{\phi\phi}}{2mR^2} \!+\! V(\phi) \!+\! \tilde{g} |\Psi_\downarrow|^2 \!
    \end{pmatrix}\!,
\end{equation}
where $R$ is the average ring radius and $\Delta = \beta(\pi/a)^2$ characterizes the TE-TM SO coupling strength enhanced by confinement (see Appendix~\ref{ssec1} for the derivation), as
shown in Fig.~\ref{fig1}(a).
The effective interaction constant $\tilde{g}$ of the 1D problem is determined from $g$ by integration with the radial wave function.
In the weakly-nonlinear limit the two lowest-lying states $\psi_{\rm g}(\phi)$ and $\psi_{\rm e}(\phi)$ of the scalar Gross-Pitaevskii equation $[- \hbar^2 \partial^2_{\phi\phi}/2mR^2  + V(\phi) + \tilde{g} N |\psi|^2]\psi = E\psi$ (the solid lines in Fig.~\ref{fig1}(c), $N$ is the particle number) are almost degenerate and substantially separated from all the higher-energy states, which allows to truncate the basis and change to the functions localized in the left ($L$) and right ($R$) parts of the ring $\Phi_{L,R}(\phi) = (\psi_{\rm g}\pm\psi_{\rm e})/\sqrt{2}$. Integrating out the coordinate dependence in the second-quantized Hamiltonian leads to the commonly-used two-mode model for the BJJ~\cite{gati2007bosonic}.

To capture the influence of the pseudospin structure of the exciton-polariton fluid on its Josephson dynamics, we generalize this approach to four modes. Due to the TE-TM splitting, spin oscillations can occur between the fractions of quasiparticles with a certain pseudospin projection. Thus, the total order parameter of the condensate can be represented as a superposition of four wave functions with time-dependent complex coefficients $\psi_{i\sigma}(t)$ describing the number of particles $N_{i\sigma}(t) = |\psi_{i\sigma}(t)|^2$ with $i = L,R$ and the spin projection $\sigma=\uparrow,\downarrow$.

Starting with the Hamiltonian (\ref{hamiltonian_spinor}) we obtain the set of equations (the full derivation is provided in Appendix~\ref{ssec3}):
\begin{multline}\label{seteq}
    i \hbar \partial_t\psi_{i\sigma} \!=\! E_0^\sigma\psi_{i\sigma} + \bigl(2U_{\!\sigma} |\psi_{i\sigma}|^2 \!-\! K_1^\sigma \psi_{\bar{i}\sigma}^*\psi_{i\sigma}\bigr)\psi_{i\sigma}  \\
    - \bigl(K_0^\sigma + 2 K_1^\sigma |\psi_{i\sigma}|^2 + K_1^\sigma |\psi_{\bar{i}\sigma}|^2\bigr)\psi_{\bar{i}\sigma} + J_{\sigma\bar{\sigma}} \psi_{i\bar{\sigma}},
\end{multline}
where the bar above any index denotes the opposite species ($\bar{L}=R$, $\bar{\uparrow}=\,\downarrow$, etc.), while the energy offsets
\begin{equation} \nonumber
    E_0^\sigma\!, K_0^\sigma \!=\! \frac12 \! \int_0^{2\pi} \!\!\!\!\! d\phi \left\{\psi_{\rm e}\!\!\left[\!-\tfrac{\hbar^2\psi_{\rm e}^{\prime\prime}}{2mR^2} \!+\! V\psi_{\rm e}\!\right] \!\pm\!  \psi_{\rm g} \!\!\left[\!-\tfrac{\hbar^2\psi_{\rm g}^{\prime\prime}}{2mR^2} \!+\! V\psi_{\rm g}\!\right] \! \right\}\!
\end{equation}
and the coefficients governing the dynamics of Eqs.~\eqref{seteq}
\begin{equation}\label{UKJ}
    \begin{split}
        & U_{\!\sigma} = \tilde{g} \!\!\int_{0}^{2\pi} \!\!\!\!\!\! d\phi \, |\psi_{\rm g}|^2 |\psi_{\rm e}|^2\!, \,\, K_1^\sigma = \frac{\tilde g}{4} \! \int_{0}^{2\pi}\!\!\!\!\!\! d\phi \, |(\psi_{\rm e}|^4 \!-\! |\psi_{\rm g}|^4), \\
        & J_{\uparrow \downarrow} = J_{\downarrow \uparrow}^* = \frac{\Delta}{2} \int_{0}^{2\pi} \!\!\!\!d\phi \left(|\psi_{\rm e}|^2 + |\psi_{\rm g}|^2\right) e^{2i\phi}
    \end{split}
\end{equation}
are defined from the two lowest-lying states  $\psi_{\rm g}$ and $\psi_{\rm e}$ of the 1D problem [see Fig.~\ref{fig1}(c)] solved for a given number of particles $N_\sigma = |\psi_{L\sigma}|^2 + |\psi_{R\sigma}|^2$ with a pseudospin projection $\sigma$, and thus they are density-dependent. We note that the ring geometry that we consider provides the controllability of the spinor junction with respect to the spin-flip rate $J_{\uparrow\downarrow}$, both by tuning the TE-TM splitting strength via the ring width $a$ and by shifting the position of the barrier junctions along the ring, which would lead to the appearance of the imaginary part of $J_{\uparrow\downarrow}$. Generally, considering the four-mode model~\eqref{seteq}, a microcavity with a linear-polarization dichroism, which can be induced and controlled by strain or applied electric field, would lead to the same dynamical equations, even though with a different SO-coupled Hamiltonian compared to~\eqref{hamiltonian_spinor}. The influence of dissipation, which can be phenomenologically described by replacing $E_0^\sigma \to E_0^\sigma - i\hbar\gamma/2$, where $\gamma$ is the spin-independent polariton decay rate, is discussed below.

\section{Parameters}\label{sec_params}
To ensure the applicability of the four-mode model and the validity of the condition $\xi\ll a$, we consider the low-density limit.
In particular, we vary the average total particle density $\rho_{tot} = \rho_\uparrow+\rho_\downarrow$ from 50 to 200~$\mu$m$^{-2}$. Within this range, direct calculation of the overlap integrals~\eqref{UKJ} allows to take approximately $U_{\uparrow,\downarrow}=U$, $K_1^{\uparrow,\downarrow}=K_1$ and $J_{\uparrow\downarrow}\approx J_{\downarrow\uparrow}=J$ (see Tables~\ref{tab2},\ref{tab3} and Fig.~\ref{fig_density} in Appendix). Given that the photon-exciton detuning in polariton systems allows to tune the polariton interactions, the average effective mass, and the ratio $m_{\rm TM}/m_{\rm TE}$ in a wide range of values, we assume working in the regime of negative detunings which lead to weaker interactions but a higher variability of the TE-TM parameter $\Delta$. This weakly-interacting regime provides conditions for the most interesting dynamics of the SO-coupled BJJ. We vary the 2D interaction constant of polaritons with the same spin $g$ from $0.2$ to $1~\mu$eV~$\mu$m$^2$ (neglecting the interaction of polaritons with opposite spins) and the magnitude of the TE-TM splitting in terms of $\Delta$ from 0.005 to 0.02~meV, which can be achieved by changing the ring width $a$ at a fixed $\beta \simeq 8~\mu$eV~$\mu$m$^2$ [the masses ratio $m_{\rm TM}/m_{\rm TE} = 0.998$, see Fig.~\ref{fig1}(a)].

We note that for the barrier to act as a weak link, the width of the barrier $\alpha R$ should be comparable with the healing length $\xi$. If $\alpha R<\xi$, the condensate order parameter does not get suppressed even when the barrier is high (we vary the barrier heights $V$ for relevant $\alpha$; for thin barriers, at $V\to\infty$ one arrives at $\delta$-functional defects). Therefore the lower boundary for $\alpha$ is roughly defined by the healing length at a given average density. On the other hand, if the barriers are too wide, they fill the available space on the ring, and the energy levels are pushed out of the potential wells. This defines the upper boundary for $\alpha$: at least two discrete energy levels should exist and they should be separated from the continuous part of the spectrum. The extended analysis of lowest-lying wavefunctions dependent on the barrier geometry is provided in Appendix~\ref{ssec1} (in particular, see Fig.~\ref{fig:notun}).

The values characterizing the Josephson dynamics are the particle imbalance between the left and right half-rings $z(t)=[N_{\!L}(t)-N_{\!R}(t)]/N_{tot}$ (regardless of spin) and the DCP of the fluid $\wp_c^{L(R)}(t) = [N_{\!L(R)\uparrow}(t) - N_{\!L(R)\downarrow}(t)]/N_{\!L(R)}(t)$ in each half-ring, with $N_{\!L(R)}$ denoting the number of particles on the left (right) half and $N_{tot}$ is the total particle number.
The initial DCP is assumed the same for the whole ring, i.e. $\wp_c^L(0)=\wp_c^R(0)=\wp_c(0)$.
Such a configuration is most easily controlled experimentally. More precisely, we keep in mind the situation when the system is excited with a Gaussian beam with a given $\wp_c$; at the same time, shifting the position of the excitation spot with respect to the ring center would allow to change the
particle imbalance $z$ on the two sides of the ring at will. Therefore, for each selection of \textcolor{NavyBlue}{$V$, $\alpha$,} $\rho_{tot}$, $g$, and $\Delta$, our control parameters are $z(0)$ and $\wp_c(0)$ which determine the initial values $\psi_{i\sigma}(0)$.

\vspace{-3pt}
\section{Results} Solving the set of Eqs.~\eqref{seteq} numerically, we follow the dynamics of the extrinsic and intrinsic Josephson effects [i.e. the evolution of $z(t)$ and $\wp_c^{L(R)}(t)$]. In general, their dynamics cannot be analyzed separately except for the cases when only one (extrinsic or intrinsic) Josephson effect is maintained while the other one is absent.
To get some analytical insight, we plug $\psi_{i\sigma}(t)=\sqrt{N_{i\sigma}(t)}e^{i\delta_{i\sigma}(t)}$ into Eqs.~\eqref{seteq}. Separating the real and imaginary parts and deducing the quantities of interest, we get:
\begin{widetext}
\begin{align}
    \dot{z}_\uparrow & = -\frac{2\Omega_\uparrow}{\hbar} \sqrt{1-z_\uparrow^2} \sin\delta_\uparrow + \frac{J}{\hbar}\left[\frac{N_L}{N_\uparrow}(1-z_\uparrow)\sqrt{1-(\wp_c^L)^2}\sin\delta^L_{\downarrow\uparrow} - \frac{N_R}{N_\uparrow}(1 + z_\uparrow)\sqrt{1-(\wp_c^R)^2} \sin\delta^R_{\downarrow\uparrow}\right], \label{ext1} \\
    \dot{z}_\downarrow & = -\frac{2\Omega_\downarrow}{\hbar}\sqrt{1-z_\downarrow^2}\sin\delta_\downarrow - \frac{J}{\hbar}\left[\frac{N_L}{N_\downarrow}(1-z_\downarrow)\sqrt{1-(\wp_c^L)^2}\sin\delta^L_{\downarrow\uparrow} - \frac{N_R}{N_\downarrow}(1 + z_\downarrow)\sqrt{1-(\wp_c^R)^2}\sin\delta^R_{\downarrow\uparrow}\right], \label{ext2}
    \\
   \dot{\delta}_{\sigma} & = \frac{2U_\sigma N_{\sigma}}{\hbar}z_{\sigma} + \frac{2 \Omega_{\sigma}}{\hbar} \frac{z_{\sigma}}{\sqrt{1-z_{\sigma}^2}}\cos \delta_{\sigma} - \frac{2J}{\hbar} \left(\frac{N_{R\bar{\sigma}}}{N_R} \frac{ \cos\delta_{\downarrow\uparrow}^R}{\sqrt{1 - (\wp_c^R)^2}} - \frac{N_{L\bar{\sigma}}}{N_L} \frac{ \cos\delta_{\downarrow\uparrow}^L}{\sqrt{1 - (\wp_c^L)^2}}\right)\!, \label{ext3}
    \\
    \dot{\wp}_c^L & = - (1-\wp_c^L)\frac{\Omega_\uparrow }{\hbar} \frac{N_\uparrow}{N_L} \sqrt{1-z_\uparrow^2} \sin\delta_\uparrow + (1+\wp_c^L) \frac{\Omega_\downarrow }{\hbar} \frac{N_\downarrow}{N_L}\sqrt{1-z_\downarrow^2} \sin\delta_\downarrow + \frac{2J}{\hbar} \sqrt{1 \!-\! (\wp_c^L)^2} \sin\delta_{\downarrow\uparrow}^L, \label{int1}
    \\[8pt]
   \dot{\wp}_c^R & = (1-\wp_c^R)\frac{\Omega_\uparrow }{\hbar} \frac{N_\uparrow}{N_R} \sqrt{1-z_\uparrow^2} \sin\delta_\uparrow - (1 +\wp_c^R) \frac{\Omega_\downarrow }{\hbar} \frac{N_\downarrow}{N_R} \sqrt{1-z_\downarrow^2} \sin\delta_\downarrow + \frac{2J}{\hbar}  \sqrt{1 \!-\! (\wp_c^R)^2} \sin\delta_{\downarrow\uparrow}^R, \label{int2}
    \\[8pt]
    \dot{\delta}_{\downarrow\uparrow}^{i} & = \frac{E_0^\uparrow \!-\! E_0^\downarrow}{\hbar} \!+ \frac{2(U_{\!\uparrow} N_{i\uparrow} \!-\! U_{\!\downarrow} N_{i\downarrow}\!)}{\hbar}
    \!- \frac{2J}{\hbar}\!\frac{\wp_c^{i}}{\sqrt{1 \!-\! (\wp_c^{i})^2}} \cos\delta_{\downarrow\uparrow}^{i} \!+ \frac{\Omega_\downarrow \!+\! 2K_1^\downarrow N_{i\downarrow}}{\hbar} \! \sqrt{\!\frac{N_{\bar{i}\downarrow}}{N_{i\downarrow}}} \cos\delta_{\downarrow}
    \!-\! \frac{\Omega_\uparrow \!+\! 2K_1^\uparrow N_{i\uparrow}}{\hbar} \! \sqrt{\!\frac{N_{\bar{i}\uparrow}}{N_{i\uparrow}}} \cos\delta_{\uparrow},  \label{int3}
\end{align}
\end{widetext}
where $z_{\sigma} = (N_{L\sigma} - N_{R\sigma})/N_{\sigma}$, $\delta_{\sigma} = \delta_{R\sigma} - \delta_{L\sigma}$ define the extrinsic Josephson dynamics in each pseudospin component, $\Omega_\sigma = K_0^\sigma+K_1^\sigma N_\sigma$, and $\delta_{\downarrow\uparrow}^{i} = \delta_{i\downarrow} - \delta_{i\uparrow}$.

\vspace{-5pt}
\subsection{Spatial oscillatory dynamics}

Equations~\eqref{ext1}--\eqref{ext3} describe the spatial (extrinsic) Josephson dynamics. Since the overall phase of the left and right wells of the BJJ cannot be properly defined anymore, we choose to track only the evolution of the overall particle imbalance $z(t)$:
\begin{equation}\label{total}
\dot{z}  \!=\! -\frac{2\Omega_\uparrow}{\hbar}\frac{N_\uparrow}{N_{tot}}\sqrt{1 \!-\! z_\uparrow^2}\sin\delta_\uparrow - \frac{2\Omega_\downarrow}{\hbar}\frac{N_\downarrow}{N_{tot}}\sqrt{1 \!-\! z_\downarrow^2} \sin\delta_\downarrow.
\end{equation}
\begin{figure}[b!]
    \centering
    \includegraphics[width=\linewidth]{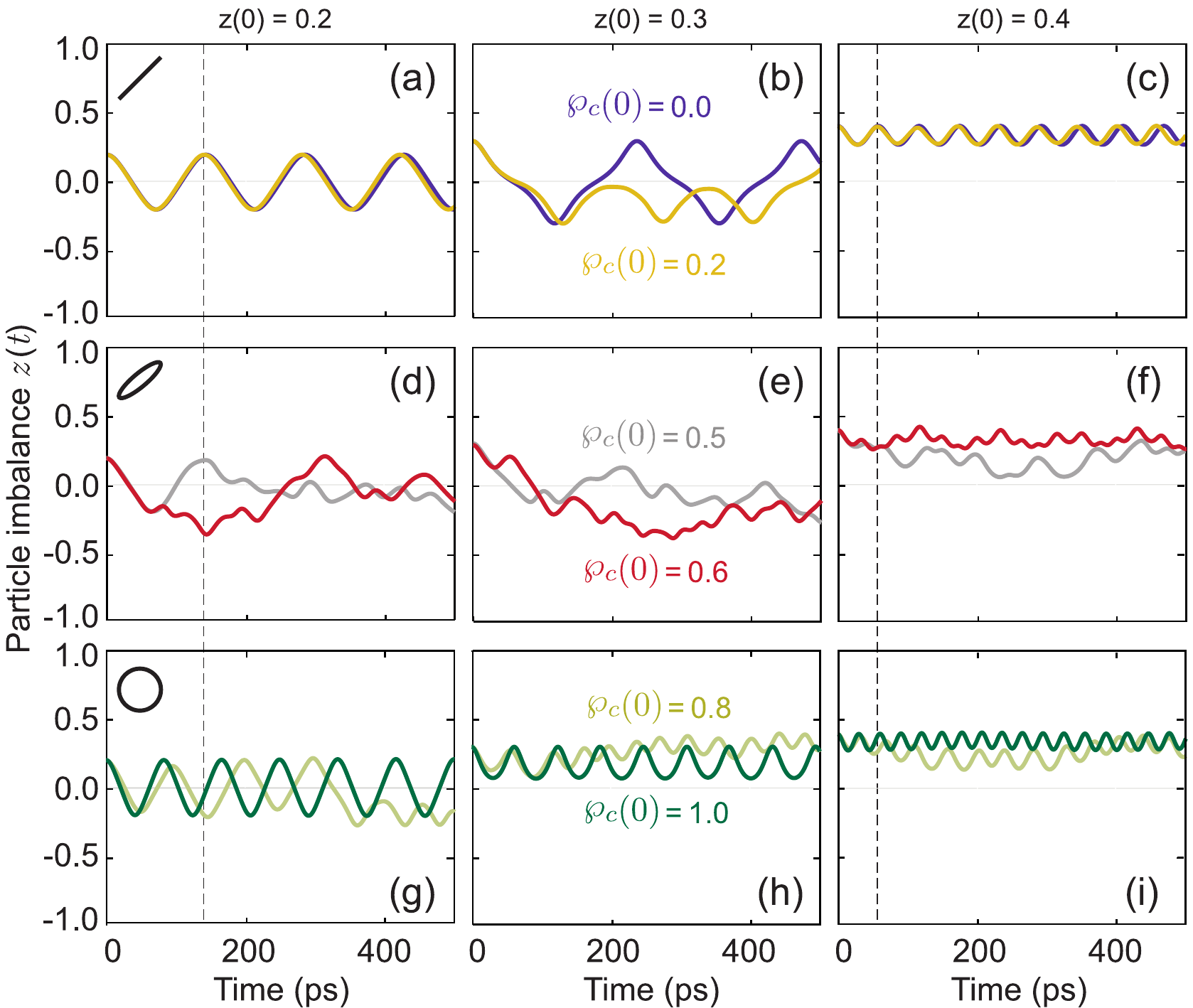}
    \caption{{\bf The spatial Josephson effect in presence of SO coupling.} Evolution of the particle imbalance obtained from the four-mode dynamical model~\eqref{seteq} for different initial values $z(0)$: (a,d,g) $0.2$, (b,e,h) $0.3$, (c,f,i) $0.4$ and various initial DCP of the fluid. (a--c) Linear-polarization case $\wp_c(0)\to 0$ resembles the scalar-case result.
    (d--f) As $\wp_c(0)$ approaches $\wp_c^{\rm crit}$, the spatial Josephson oscillations are destroyed.
    (g--i) When $\wp_c(0)>\wp_c^{\rm crit}$ and approaches 1 (circular polarization), the conventional regimes are recovered with the doubled particle number. Vertical dashed lines are guides to the eye for comparison of oscillation frequencies.
    Parameters: $m = 10^{-5}m_0$, $\Delta=0.02$~meV, $g = 1~\mu$eV~$\mu$m$^2$, $\rho_{tot} = 100~\mu$m$^{-2}$, $V=0.5$~meV, $\alpha = 0.8$.}
    \label{fig2}
\end{figure}
For extrinsic Josephson effect, the truncation of the SO-coupled BJJ to a conventional double-well situation is realized if $\wp_c(0)=0$ (linearly-polarized condensate) or $\wp_c(0) = \pm1$ (purely circular polarization). The critical value of $z(0)$ to pass to the MQST regime is
\begin{equation}\label{Zcrit}
    z^{\rm crit} = \pm \frac{2 (K_0^\sigma + K_1 N_\sigma)}{U N_\sigma} \sqrt{\frac{U N_\sigma}{K_0^\sigma + K_1 N_\sigma} - 1},
\end{equation}
where $N_\sigma = N_{tot}/2$ for linear and $N_\sigma = N_{tot}$ for circular polarization. These specific cases are shown in Fig.~\ref{fig2}, panels (a--c) and (g--i), respectively. The oscillation frequency for a circularly-polarized BJJ is almost twice higher due to the doubled particle number in one pseudospin component. For the chosen parameters, the critical values $z_{\rm lin}^{\rm crit} = 0.306$ and $z_{\rm circ}^{\rm crit}=0.293$, hence the MQST regime for circularly-polarized condensate is reached sooner [cf. panels (b) and (h) in Fig.~\ref{fig2}, both for $z(0)=0.3$]. When the polarization is slightly different from linear or circular, the spatial dynamics is qualitatively unchanged: at small $z(0)$ it performs sinusoidal oscillations which become anharmonic when $z(0)$ approaches $z^{\rm crit}$, and at $z(0)>z^{\rm crit}$ the junction is in the MQST regime. As $\wp_c(0)$ starts to deviate from 0 or 1, the frequency of oscillations shifts and the amplitude is slightly reduced in all regimes [see Fig.~\ref{fig2}(a--c) and (g--e)]. Notably, when $\wp_c(0)$ strongly deviates from 0 or 1 (elliptical polarization) and is in the vicinity of $\wp_c^{\rm crit}$, the spatial oscillations are destroyed [see Fig.~\ref{fig2}(d--f)] by the strong influence of the critical behavior in the condensate polarization dynamics.

\vspace{-5pt}
\subsection{Fixed points}

Similarly to the standard two-mode Josephson dynamics (see Appendix~\ref{ssec2}), the fixed points of Eq.~\eqref{total} correspond to the zero- and $\pi$-phase modes of the inter-well phase difference, i.e. $\delta_\sigma\in\{0,\pi\}$. For the considered experimental configuration, however, only zero-phase modes are relevant. With $\sin\delta_\sigma = 0$, the equations~\eqref{ext1}--\eqref{ext2} for $\dot{z}_\sigma$ and \eqref{int1}--\eqref{int2} for $\dot{\wp}_c^{L,R}$ simplify and yield the stationary states that must satisfy either $|\wp_c^i|=1$ (corresponding to a fully circularly-polarized condensate on the $i$-th side of the ring), or $\sin\delta_{\downarrow\uparrow}^i = 0$ if $|\wp_c^i|<1$. The remaining stationarity conditions $\dot{\delta}_\sigma=0$ and $\dot{\delta}_{\downarrow\uparrow}^i=0$ then reduce to a set of coupled algebraic relations between $z_\uparrow$, $z_\downarrow$, $\wp_c^L$, $\wp_c^R$
and the system parameters $U_\sigma$, $J$, $\Omega_\sigma$, $K_1^\sigma$ and $E_0^\uparrow - E_0^\downarrow$. While this nonlinear algebraic system cannot be solved in closed form in full generality, it is evident that physically relevant classes of stationary solutions include the symmetric configuration
$\langle z_\uparrow\rangle = \langle z_\downarrow\rangle =0$, $\langle\wp_c^L\rangle =
\langle\wp_c^R\rangle=0$, $\langle\delta_\uparrow\rangle = \langle\delta_\downarrow\rangle = \langle \delta^L_{\downarrow\uparrow}\rangle = \langle \delta^R_{\downarrow\uparrow}\rangle =0$ [see Fig.~2(a--b)], spatially self-trapped solutions with $\langle z_\sigma\rangle\neq 0$, corresponding to an imbalance of population between the two traps that is dynamically sustained by the nonlinearity [Fig.~2(c), (h--i)], and the spin-polarized stationary states with  $\langle\wp_c^i\rangle\neq0$ (we will refer to this regime as polarization self-localization). Such states with non-zero mean circular polarization can coexist both with the zero-phase oscillations of $z_\sigma$ and with coordinate self-trapping, or, on the contrary, hinder the spatial oscillations. Overall this interplay leads to a rich set of polarization regimes in the full four-mode dynamics.

\vspace{-5pt}
\subsection{Polarization oscillations and switching}

\begin{figure}[b!]
\centering\includegraphics[width=1.\linewidth]{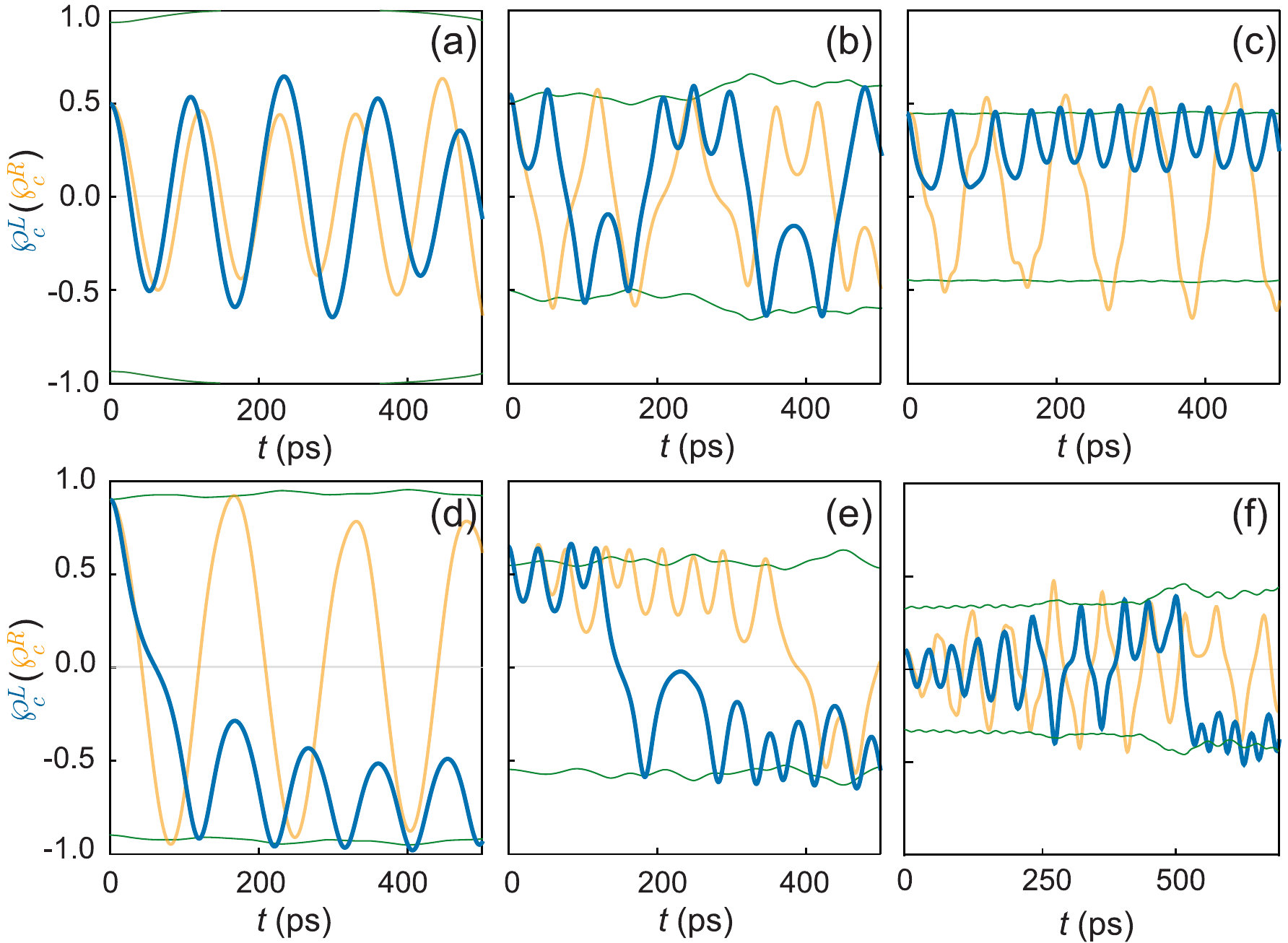}
\caption{\textbf{Polarization oscillations and polarization switching.} (a) Evolution of the DCP on the left (the blue lines) and right (the yellow lines) half-rings without switching (top) and in the switching regime (bottom). Parameters: $m=10^{-5}m_0$, $\Delta = 0.02$~meV; $\rho_{tot} = 100$ (a--e) and $200~\mu$m$^{-2}$ (f); $g=0.2$ (a,d) and $1~\mu$eV~$\mu$m$^2$ (b,c,e,f). The following regimes are displayed: (a) conventional harmonic oscillations [$z(0) = 0.4$, $\wp_c(0) = 0.5$]; (b) anharmonic anti-phase oscillations of the DCP of the left and right half-rings [$z(0) = 0.25$, $\wp_c(0) = 0.55$]; (c) self-localization of $\wp_c^L$, accompanied by the regular oscillations of $\wp_c^R$ [$z(0) = 0.65$, $\wp_c(0) = 0.45$]; (d) polarization switching from anharmonic oscillations [$z(0) = 0.6$, $\wp_c(0) = 0.9$]; (e) polarization switching from self-trapped oscillations [$z(0) = 0.05$, $\wp_c(0) = 0.65$]; (f) the DCP of the left half-ring undergoes all regimes in sequence, harmonic--anharmonic--self-trapped and then switches sign [$z(0) = 0.45$, $\wp_c(0) = 0.1$]. The green lines show the critical value $\wp_{c\,L}^{\rm crit}(t)$.
}
\label{fig3}
\end{figure}

\begin{figure}[t!]
    \includegraphics[width=\linewidth]{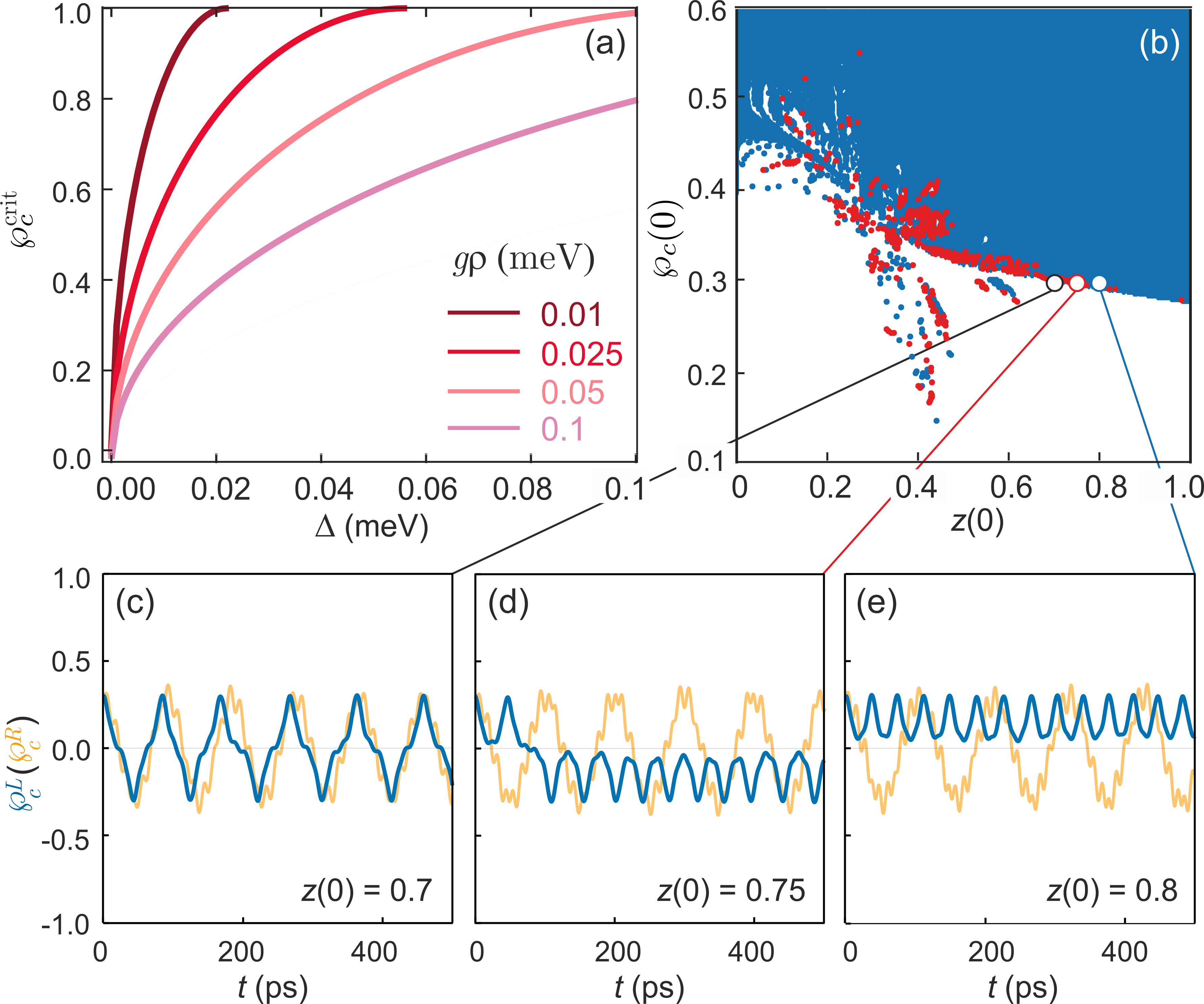}
\caption{\textbf{Diagram of polarization oscillations regimes.}
(a) Dependence of the critical value $\wp_c^{\rm crit}$ of the DCP defining the transition to the polarization self-trapping regime versus the TE-TM parameter $\Delta$, in the absence of the external Josephson effect [$z(0)=0$], for different nonlinearity values.
(b) Diagram of dynamical regimes of the DCP oscillations depending on the initial values $\wp_c(0)$ and $z(0)$ for the left half-ring. White: oscillations with zero average, blue: self-localization of polarization, red: the switching regime.
(c--e) Anharmonic oscillations (c), polarization switching (d), and the self-trapping regime (e) realized across the critical region [see panel (b)] for $\wp_c(0)=0.3$.
In panels (b--e), $\Delta = 0.02$~meV, $g = 1~\mu$eV~$\mu$m$^2$, $\rho_{tot} = 200~\mu$m$^{-2}$. 
}
\label{fig4}
\end{figure}

The critical value of the DCP for polarization self-localization is easily calculated for the case $z(0)=0$:
\begin{equation}\label{Pcrit}
    \wp^{\rm crit}_{c} = \frac{4 J}{(U-2K_1)N_{tot}}\sqrt{\frac{U-2K_1}{2J}N_{tot}-1},
\end{equation}
and is shown in Fig.~\ref{fig4}(a) against the TE-TM splitting parameter $\Delta$. When $z(0)$ is nonzero, $\wp_c^{\rm crit}$ is time-dependent and is generally different for the left and right halves of the ring. In this case, $\wp_{c\,L(R)}^{\rm crit}(t)$ can be obtained from Eq.~\eqref{Pcrit} approximately, by replacing $N_{tot}/2 \to N_{L(R)}(t)$. This is the weak variation of $\wp_{c\,L,R}^{\rm crit}$ in time that results in the novel dynamical regimes in the behavior of DCP of the two half-rings (intrinsic Josephson effect).

In the absence of the initial population imbalance $z(0)=0$, the number of particles in each half-ring is preserved ($N_L=N_R$ and the effective tunneling is zero), so both $\wp_c^L$ and $\wp_c^R$ exhibit coinciding oscillations in the conventional BJJ regimes~\cite{gati2007bosonic} with the energy detuning between the pseudospin components (see Eqs.~\eqref{internal} in Appendix~\ref{ssec3}).
However, when $z(0)$ is varied, the system demonstrates more complex and diverse dynamics. Differing initial particle numbers $N_{L,R}(0)$ result in differing critical values $\wp_{c\, L,R}^{\rm crit}$, and, due to the flow of particles between half-rings, these critical values are changing in time. The top row of Fig.~\ref{fig3} shows three distinct regimes of internal Josephson effect with SO coupling accompanied by the oscillations of $z(t)$. In (a), the DCP in both left and right half-rings experience conventional sinusoidal oscillations ($\wp_c^{L,R}<\wp_c^{\rm crit}$ at all times). In (b), $\wp_c(0)$ is slightly supercritical for the left half, while being subcritical for the right. The variation of $\wp_{c\,L}^{\rm crit}$ in time results in anti-phase nonlinear oscillations of $\wp_c^L(t)$ and $\wp_c^R(t)$. In (c), $N_L$ largely exceeds $N_R$ during the whole evolution. Since the critical value $\wp_c^{\rm crit}$ rapidly drops with the growth of density, $\wp_c^L$ is self-localized, while $\wp_c^R$ is oscillating around zero.

Interestingly, the fine-tuning of the initial DCP value $\wp_c(0)$ in the vicinity of $\wp_c^{\rm crit}$ and the difference of the critical values $\wp_{c\,L}^{\rm crit}\neq\wp_{c\,R}^{\rm crit}$ allows to achieve the `polarization switching' regime which was not previously reported. In this parameter range, the spatial dynamics of populations is not oscillatory [see Fig.~\ref{fig2}(d)--(f)], and the particle numbers $N_{L,R}$ (together with $\wp_{c\,L,R}^{\rm crit}$) only slightly drift in time.
If during the evolution $\wp^{\rm crit}_c$ for the $i$-th half-ring decreases or increases, $\wp_c^i(t)$ may switch from the oscillatory to the self-trapped dynamics and vice versa; that happening at specific moments of time leads to the change of sign of the DCP to the opposite [see Fig.~\ref{fig3}(d,e)]. Depending on parameters, the DCP of the opposite half-ring may be oscillating (d) or stay self-trapped for a longer time (e). Finally, Fig.~\ref{fig3}(f) shows a realization in which $\wp_c^L(t)$ passes through all possible Josephson regimes (harmonic and anharmonic oscillations, self-localization, and polarization switching). We note that in the parameter range corresponding to Fig.~\ref{fig3}(e,f), the time traces of the irregular switching of the DCPs $\wp_c^{R,L}(t)$ do not settle into a simple periodic or quasiperiodic pattern. This behavior is compatible with the onset of deterministic chaos, and the range of the initial conditions identified that lead to this weakly chaotic behavior is presented by the red dots in the diagram of Fig.~\ref{fig4}(b). Overall, Fig.~\ref{fig4}(b) summarizes the parameter values at which various regimes are achieved, and indicates more specific cases realized when $z(0)$ is tuned at a fixed $\wp_c(0)$ in the critical region [see Fig.~\ref{fig4}(c--e)]. By choosing $z(0)$, one can achieve a subtle condition where self-localization results in the DCP inversion on one or both halves of the ring. More examples of various regimes and parameter diagrams for varying nonlinearity, TE-TM splitting, and geometry of the barriers are presented below.

\vspace{-5pt}
\subsection{Robustness against the geometry change}

In the results reported above, we have not varied the geometry of the system, but tracked various regimes of oscillations depending on the initial conditions $z(0)$, $\wp_c(0)$ and the system's nonlinearity $g\rho$. In particular, for Figures~\ref{fig2}--\ref{fig4} [except~\ref{fig4}(a)], we took the ring thickness $a=2~\mu$m, which yields $\Delta = 0.02$~meV [see the red star in Fig.~\ref{fig1}(a)], and the barriers parameters (height and the angular width) were fixed at $V=0.5$~meV, $\alpha=0.8$. In this subsection, we address the accessibility of all the reported regimes when the barriers geometry is changed.

\begin{figure}[b!]
    \centering
    \includegraphics[width=\columnwidth]{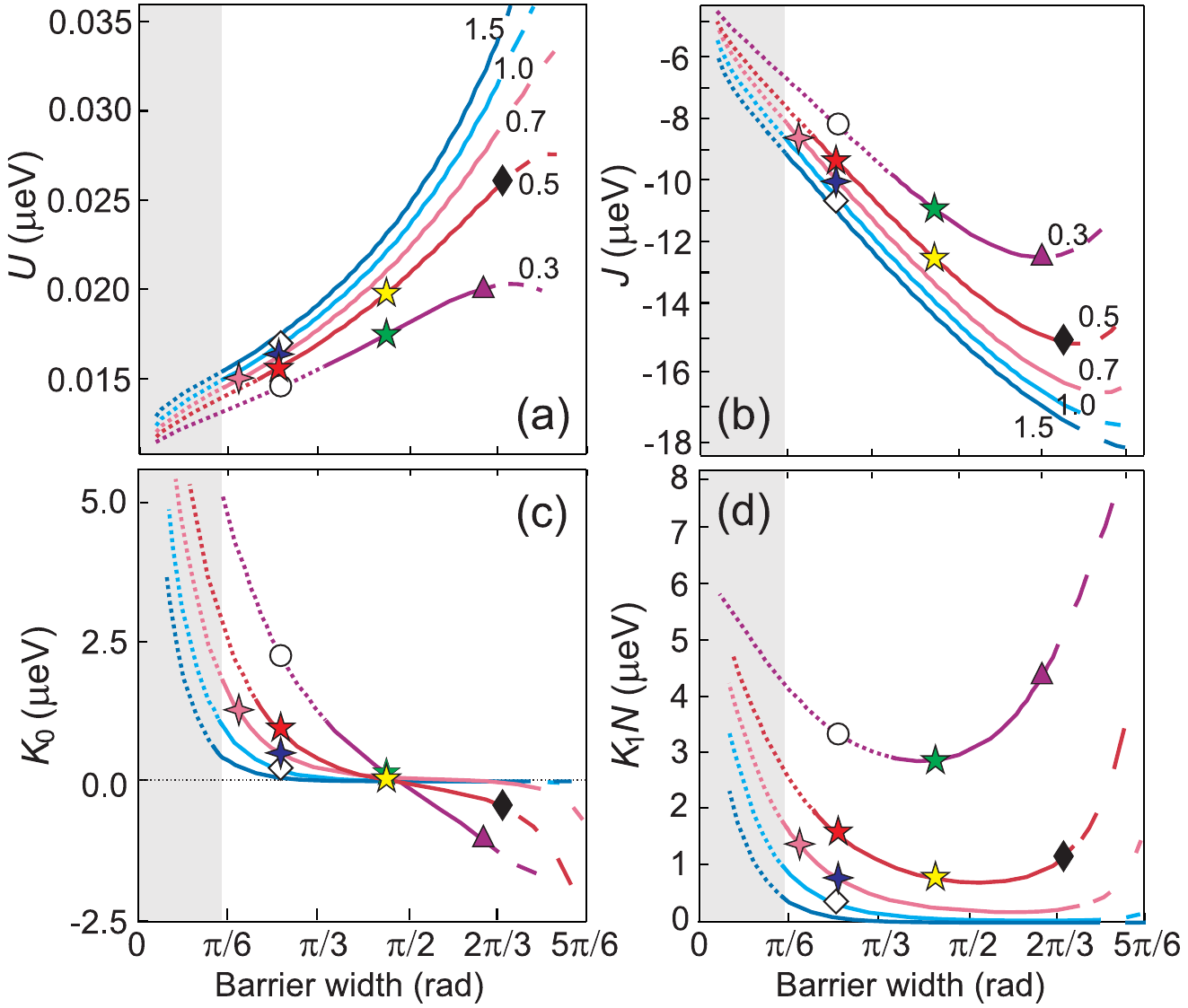}
    \caption{{\bf Parameters governing the dynamics versus the barriers width $\alpha$}. According to Eq.~\eqref{UKJ}, we plot the interaction parameter $U$ (a); the spin-flip rate $J$ (b); the parameters defining the effective tunneling rate $K_0$ (c) and $K_1N$ (d). In all panels, the full condensate density $\rho_{tot} = 100~\mu$m$^{-2}$ and $g=1~\mu$eV~$\mu$m$^2$, $\Delta = 0.02$~meV. Different lines correspond to different barriers height $V$, as marked in (a--b): 0.3~meV (purple), 0.5~meV (red), 0.7~meV (pink), 1.0~meV (light blue), and 1.5~meV (dark blue). The gray-shaded region marks the widths smaller than the healing length. The dotted (dashed) parts of the lines mark the regions of widths where the barriers are too narrow (too wide) to work as Josephson junctions. Various markers (circles, diamonds, stars, triangles) mark the parameters used in simulations (see Fig.~\ref{fig6} and Fig.~\ref{fig:notun}).}
    \label{fig5}
\end{figure}

As discussed in Sec.~\ref{sec_params} and as can be anticipated from Fig.~\ref{fig1}(c), not every combination of the barrier width and height provides conditions for the Josephson tunneling. In Fig.~\ref{fig5} we investigate the behavior of the overlap integrals~\eqref{UKJ} governing the dynamics on the barrier height and width, at a fixed density $\rho_{tot}$ and interaction constant $g$. The irrelevant region $\alpha R<\xi$ is shaded gray; furthermore, by plotting the ground- and excited-states wavefunctions of the GPE (see Fig.~\ref{fig:notun} in Appendix~\ref{ssec1}), we find that depending on the barrier height $V$, the region of the widths where the barriers are acting as tunnel junctions is even narrower. The dotted part of the lines in Fig.~\ref{fig5} marks the widths at which the order parameter in the under-barrier region is not suppressed, and hence there's no defined spatial separation between the left and right parts of the ring. The dashed parts, on the other hand, denote the region where the barriers are too wide and the energy levels are too close to or above the top of the barriers.

\begin{figure}[b!]
    \centering
    \includegraphics[width=\columnwidth]{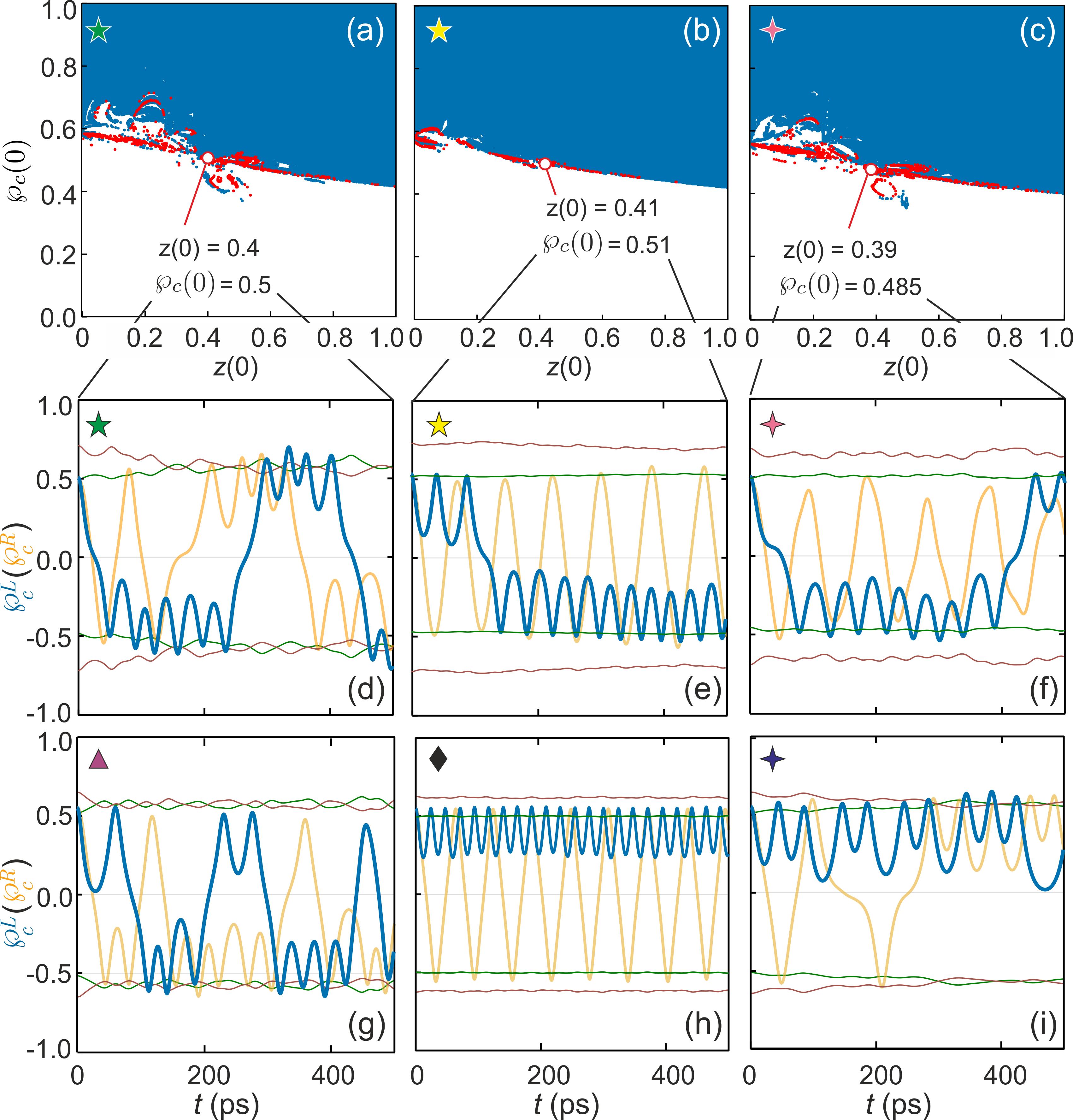}
    \caption{{\bf Realizations of DCP oscillation regimes while varying geometry of the barriers.} The three columns correspond to different barrier heights $V=0.7$~meV (left), $0.5$~meV (middle) and $0.3$~meV (right). (a,b,d,e) $\alpha = 1.35$; (c,f) $\alpha = 0.6$; (g) $\alpha = 2.0$; (h) $\alpha=2.2$; (i) $\alpha = 0.8$. The stars, diamond and triangle in the panels correspond to the same markers in Fig.~\ref{fig5}. (a)--(c) Diagrams of initial conditions for the left half-ring, with the same color code as in Fig.~\ref{fig4}(b). (d)--(f) Examples of the polarization switching regime realized for the combinations $(V,\alpha)$ as in (a)--(c), with the initial conditions given in the panels. (g)--(i) The barrier width is taken larger (at a fixed $V$, by column), and the initial conditions (g,i) $z(0)=0.25$, $\wp_c(0)=0.55$; (h) $z(0)=0.65$, $\wp_c(0)=0.45$.}
    \label{fig6}
\end{figure}

It can be seen that the increase of the barrier heights leads to a quick decrease of the tunneling rates that are defined as $\Omega_\sigma = K_0^\sigma + K_1^\sigma N_\sigma$: while $V=0.3$~meV for most of the widths is too weak to provide substantial wavefunctions suppression, at $V$ larger than 0.7~meV basically no strong tunneling is observed [see the blue lines in Fig.~\ref{fig5}(c,d) that correspond to $V=1.0$ and 1.5~meV]. At the same time, we note that for smaller values of $V$, the parameter $K_0$ becomes negative with the growth of $\alpha$ [Fig.~\ref{fig5}(c)], which also leads to quenching of the tunneling rates for $\alpha>\pi/2$. Without the tunneling, there is no interplay between the internal and spatial Josephson effects, and the variety of the DCP oscillations regimes reported in Fig.~\ref{fig3} is reduced. Hence, we find that ``good'' weak links (for $\rho_{tot} = 100~\mu$m$^{-2}$, $g=1~\mu$eV~$\mu$m$^2$) are represented by the barriers of the height $V$ roughly in the range $[0.3,0.7]$~meV and width $\alpha\in[\pi/6,\pi/2]$. In Fig.~\ref{fig6}, we scan along different widths and heights of the barriers, and find that while staying in this range of $(V,\alpha)$, all the regimes reported previously are accessible when changing the initial conditions. The influence of the density change and the ring width variation is discussed in Appendices~\ref{ssec2} and~\ref{ssec4}.

\vspace{-5pt}
\subsection{Effect of dissipation}

Crucially, when dissipation is added phenomenologically to Eqs.~\eqref{seteq}, the shape of the Eqs.~\eqref{ext1}--\eqref{int3} that govern the evolution of the population imbalances, relative phases, and DCP on the left and right halves of the ring, does not change. More precisely, introducing the decay as imaginary parts of the energies $E_0^\sigma \to E_0^\sigma - i\hbar\gamma/2$ results in the decay of all populations $N_{i\sigma}(t)$ with the spin-independent rate $\gamma$, while the equations for the phases $\delta_\sigma^i(t)$ are mathematically not affected. As an outcome, in the Eqs.~\eqref{ext1},\eqref{ext2} for the imbalances and Eqs.~\eqref{int1},\eqref{int2} for the polarization degrees the terms containing $\gamma$ cancel out. However, the total particle number decrease $N_{tot}(t) = e^{-\gamma t}N_{tot}(0)$ leads to the decay of the nonlinearity and the shift of the critical values $z^{\rm crit}$ and $\wp_c^{\rm crit}$ with time towards higher values (see Tables~\ref{tab2} and~\ref{tab3} in Appendix). This makes the polarization switching regime reported in Fig.~\ref{fig3}(e) appear sooner in the dynamics compared to the conservative case, and a further collapse of the self-localized oscillations towards harmonic-oscillations regime shown in Fig.~\ref{fig3}(a). From the point of view of the spatial Josephson dynamics (Fig.~\ref{fig2}), the decrease of the total particle number leads to the eventual disappearance of the tunneling, since the main contribution to the tunneling rate $\Omega_\sigma = K_0^\sigma + K_1^\sigma N_\sigma$ comes from the second term [see Fig.~\ref{fig5}(e)].

\section{Conclusions}
In this work, we studied the interplay of spatial and internal Josephson dynamics for a spinor polariton condensate in ring geometry with controllable SO coupling. Depending on the condensate polarization, the tunneling dynamics of the BJJ was shown to exhibit substantially different frequencies or to totally lose the oscillatory nature. Tracking the pseudospin evolution in the four-mode model allowed to discover in-phase and anti-phase oscillations of the DCP on the left and right sides of the ring, self-trapping of polarization on one side accompanied by oscillations of the DCP around zero on the other side, and the unique switching regime not reported previously. This polarization switching regime is achievable only in a narrow range of parameters, which shrinks both with the growth of the particle density and the TE-TM splitting.

Importantly, polariton fluids are intrinsically driven-dissipative and may be subject to disorder. The pump-controlled spin switching of the SO-coupled spinor polariton fluids was studied in Ref.~\cite{PRX_Ohadi}, including that in a BJJ setting~\cite{PRL_Ohadi}. It is curious to note that the key ingredient to achieve the spin switching in those driven systems is the presence of splitting between the linear polarizations decay rates, which leads effectively to complex couplings between spin-up and spin-down components of the condensate.
In our work, on the other hand, while the ring BJJ geometry that we consider allows to achieve complex spin-flip rates, we focused on the interplay of intrinsic and extrinsic Josephson dynamics in the conservative case, without any external influence. The spin-switching that we report involves as well the switching between the ferromagnetic and antiferromagnetic configurations of polarization of the two halves of the ring, which are occurring dynamically without converging to a chosen steady state. While the effect of random disorder is expected to be integrated out in our four-mode model together with the coordinate dependence of the wave functions, the dissipation in general can not be neglected. However, we show that in the simplest possible consideration (when phenomenologically added as imaginary parts of the energies) the spin-independent finite lifetime does not change to the variety of observable regimes. The decay of the overall population with time merely brings the system to pass from one regime to the other, from the nonlinearity-dominated dynamics and towards the linear one, with gradually damped plasmalike oscillations. Hence to ensure experimental observability of the reported switching dynamics, one needs to ensure long polariton lifetime. We refer to experimental works of the past decade, including those on the ring-shaped polariton condensates~\cite{Snoke_Optica}, with particle lifetimes of 300–400~ps, allowing macroscopic flow over hundreds of microns~\cite{PRX3_2013,Optica2_2015} and equilibrium condensation~\cite{PRL118_2017,eadk6960}. The added effects of random disorder and noise are subject of a separate study.

To conclude, we note that the ability to manipulate the spin transfer, based on the interplay of the critical phenomena in the internal and external Josephson effects, is only reachable in systems with weak nonlinearity, and the high degree of control is achieved by tuning the TE-TM splitting strength by the system's geometry. The reported behavior enriches our understanding of the spinor exciton-polariton dynamics and the SO-coupled BJJs, and suggests potential applications for photonic devices and quantum information technologies.\\

\section{Acknowledgments}
The work is funded by the Russian Science Foundation grant No. 24--22--00426 (\url{https://rscf.ru/en/project/24-22-00426/}).

\begin{widetext}
\appendix

\section{One-dimensional Gross-Pitaevskii equation and the Hamiltonian~\eqref{hamiltonian_spinor}}
\label{ssec1}

In the {\bf scalar case}, the behavior of the polariton condensate wavefunction $\Psi(\mathbf{r},t)$ is governed by the two-dimensional (2D) Gross-Pitaevskii equation (GPE)
\begin{equation}\label{GPE_full}
    i\hbar \frac{\partial \Psi(\mathbf{r},t)}{\partial t} = - \frac{\hbar^2}{2m}\Delta \Psi(\mathbf{r},t) + V({\bf r})\Psi(\mathbf{r},t) + g|\Psi(\mathbf{r},t)|^2 \Psi(\mathbf{r},t),
\end{equation}
where $m$ is the polariton effective mass and $g$ the interaction constant. In the ring geometry that we consider, the external potential is represented by a sum $V({\bf r})= V(r) + V(\phi)$ of the potential in the radial direction, a ring-shaped infinite well $r_1<r<r_2$ of the width $r_2-r_1=a$, and the azimuthal potential consisting of two rectangular barriers of the height $V_{1,2}$ and angular width $\alpha$, as shown in Fig.~\ref{fig1}(b,c).

In the {\bf spinor case}, one needs to consider the two coupled GPEs for the spin-up (right-circular polarization) and spin-down (left-circular polarization) components of the polariton spinor $\Psi = (\Psi_\uparrow,\Psi_\downarrow)^T$. The  $2\times2$ Hamiltonian~\eqref{TE-TM} governing the two-component dynamics contains off-diagonal terms coming from the TE-TM splitting~\cite{shelykh2010review}:
\begin{equation}\label{hamilt_spin}
    \hat{\mathcal H} =
    \begin{pmatrix}
        -\frac{\hbar^2}{2m}\Delta + V({\bf r}) + g|\Psi_\uparrow|^2 & \beta (\partial_y + i\partial_x)^2 \\[5pt]
        \beta (\partial_y - i\partial_x)^2 & -\frac{\hbar^2}{2m}\Delta + V({\bf r}) + g|\Psi_\downarrow|^2
    \end{pmatrix},
\end{equation}
where 
we have neglected the interaction between the polaritons with opposite pseudospins.

Since we are interested in the condensate motion along the azimuthal direction, we can reduce the problem to one-dimensional (1D) assuming the thin-ring limit: $a\ll r_{1,2}$.  In the case of sufficiently thin ring (we take $r_1=10~\mu$m, $r_2=12~\mu$m, $a=2~\mu$m), the energy of quantization in the radial direction is much larger than the characteristic energy of interactions, and it is justified to separate the radial motion (temporarily dropping the nonlinear term): $\Psi(\mathbf{r},t)=\chi(r)\psi(\phi,t)$.
Following the procedure described in Refs.~\cite{gulevich2016,kozin2018topological,mukherjee2021dynamics}, we average the scalar equation~\eqref{GPE_full} with the ground-state radial wave function
$\chi_0(r) \simeq \sqrt{4/(r_2^2-r_1^2)} \sin [{\pi (r-r_1)/(r_2-r_1)}]$, which yields:
\begin{equation}
    i\hbar \frac{\partial \psi(\phi,t)}{\partial t} = - \frac{\hbar^2}{2m R^2}\frac{\partial^2}{\partial \phi^2} \psi(\phi,t) + V(\phi)\psi(\phi,t) + \tilde{g}N|\psi(\phi)|^2 \psi(\phi,t), \label{nGPE}
\end{equation}
with $R^{-2} = \int rdr|\chi_0|^2(r^{-2})$, the effective 1D interaction constant $\tilde{g} = g\int rdr |\chi_0|^4$, and $N$ being the number of particles (we choose to normalize the wave functions to unity).

Applying the same procedure to the off-diagonal terms of Eq.~\eqref{hamilt_spin} requires transition to the polar coordinates
$$(\partial_y \pm i\partial_x)^2=e^{\mp 2i\phi} \left[-\partial^2_{rr} \pm (2i/r)\partial^2_{r\phi} \mp (2i/r^2) \partial_\phi + (1/r)\partial_r + (1/r^2)\partial^2_{\phi\phi}\right],$$
which, after the averaging with $\chi_0(r)$, becomes
\vspace{-5pt}
\begin{equation}
    e^{\mp2i\phi}\left(\frac{\pi^2}{a^2} \mp \frac{2i}{R^2}\partial_\phi + \frac{1}{R^2}\partial^2_{\phi\phi}\right).
\end{equation}
Therefore for $a\ll R$, from Eq.~\eqref{hamilt_spin} one arrives at the 1D Hamiltonian~\eqref{hamiltonian_spinor} with $\Delta=\beta(\pi/a)^2$.\\

Prior to considering the two-component GPE with the Hamiltonian~\eqref{hamiltonian_spinor}, we first analyze its scalar limit. For the spinor case, this corresponds to treating one of the pseudospin components in isolation.

Our analysis and the development of the two-mode (four-mode) model relies on the assumption of a weak nonlinearity in the system, which (together with $a\ll R$) ensures the validity of truncation of Eq.~\eqref{GPE_full} to 1D, as well as using the basis $\{\psi_{\rm g}(\phi),\psi_{\rm e}(\phi)\}$ of the stationary GPE problem [see Fig.~\ref{fig1}(c)].
For the parameters used in \textcolor{NavyBlue}{most of the} simulations ($m=10^{-5}m_0$, $r_1 = 10~\mu\text{m}$, $r_2 = 12~\mu\text{m}$, $V_1 = V_2 = 0.5~\text{meV}$, $\alpha = 0.8~\text{rad}$), the numerically-found energies of the ground $E_{\rm g}$ and first excited $E_{\rm e}$ states, and the energy of the second excited state $E_3$ are listed in Table~\ref{tab1}, together with the considered interaction constants $g$ and average polariton densities $\rho$:

\begin{table}[h!]
    \centering
    \begin{tabular}{c|c|c|c|c|c|c}
        \hline
        & $g=0$ & $g = 0.2~\mu$eV~$\mu$m$^2$ &  $g = 1~\mu$eV~$\mu$m$^2$ &  $g = 1~\mu$eV~$\mu$m$^2$ &  $g = 1~\mu$eV~$\mu$m$^2$ &  $g = 1~\mu$eV~$\mu$m$^2$\\
        &  & $\rho = 50~\mu$m$^{-2}$ & $\rho = 25~\mu$m$^{-2}$ & $\rho = 50~\mu$m$^{-2}$ & $\rho = 100~\mu$m$^{-2}$ & $\rho = 200~\mu$m$^{-2}$ \\
        \hline\hline
        $\xi$ ($\mu$m) & --- & 19.5 & 12.3 & 8.7 & 6.2 & 4.4 \\
        \hline
        $E_{\rm g}$ (meV) & \qquad0.0365\qquad\, & 0.0599 & 0.093 & 0.1443 & 0.2382 & 0.4078\\
        $E_{\rm e}$ (meV) & \qquad0.0388\qquad\, & 0.0632 & 0.098 & 0.1527 & 0.255 & 0.4461\\
        $E_3$ (meV) & \qquad0.1427\qquad\, & 0.1652 & 0.198 & 0.2506 & 0.3493 & 0.5311\\
        \hline
    \end{tabular}
    \caption{Values of the 2D interaction constants and average total density used in this work, and the corresponding values of the healing length $\xi = \hbar^2/\sqrt{2mg\rho}$ and the energies of the first three stationary states of the GPE~\eqref{nGPE}.}
    \label{tab1}
\end{table}
\noindent As one can see, in the chosen geometry, for nonlinearities up to $g\rho=0.1$~meV the splitting between the first two states is from one to two orders of magnitude smaller than the gap to the second excited level.  For $\rho = 200~\mu$m$^{-2}$ (last column in Table~\ref{tab1}), this assumption is not applicable any longer. Furthermore, in the case $g\rho=0.2$~meV the healing length starts to approach the ring width $a$ and the separation of variables also loses its validity. This defines the range of the parameters used in this work.
In most of the presented results, we take the average density of one  condensate component $\rho$ from 25 to $100~\mu\text{m}^{-2}$ (for the spinor case, the total density of the two components is doubled; since the interaction between the opposite spins is negligible, the nonlinearity acts upon each pseudospin component separately). 
The polariton interaction constant is controlled by the exciton fraction of the polariton (experimentally, by the detuning between the microcavity photon mode from the exciton resonance) and can be varied in a wide range. We consider weak interactions and take $g=0.2~\mu\text{eV}~\mu\text{m}^2$ and $1~\mu\text{eV}~\mu\text{m}^2$, which lead to effective one-dimensional values $\tilde{g} \approx 0.0136~\mu$eV and $0.068~\mu$eV, respectively.
These parameters 
mark our system as the one of extremely weak nonlinearity compared to regular polariton nonlinearities in planar geometries (see e.g.~\cite{estrecho2019direct}). However we discover that even in such a dilute limit, the circular geometry allows for the observation of essentially nonlinear regimes, including all the well-known regimes of the bosonic Josephson effect (see Appendix~\ref{ssec2}). In particular, the find that the macroscopic quantum self-trapping (MQST) regime on the ring can be achieved at much smaller densities and interaction constants than in double-well polariton condensates~\cite{lagoudakis2010coherent,abbarchi2013macroscopic}. \\

\begin{figure*}[t!]
    \includegraphics[width=\textwidth]{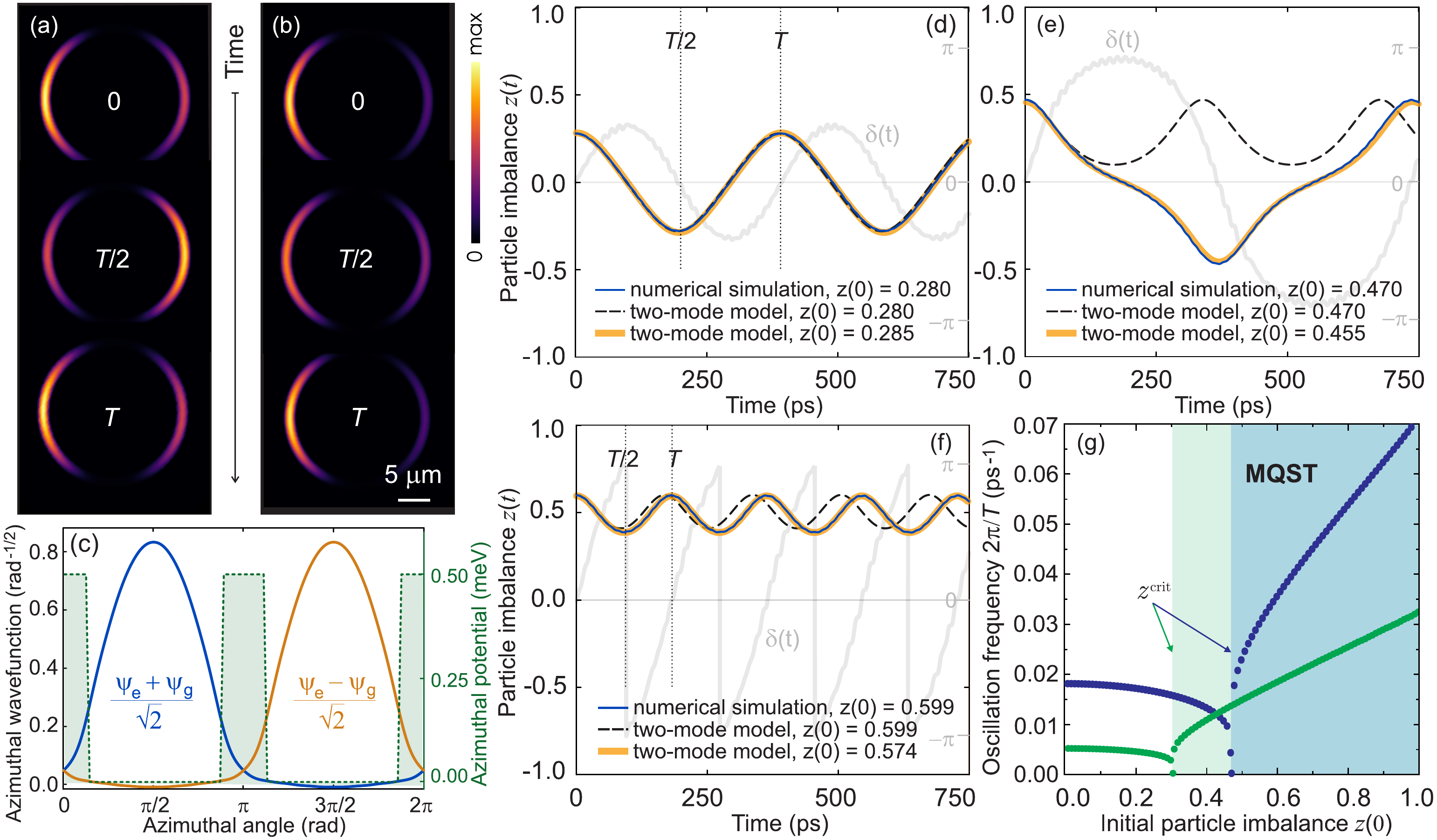}
    \caption{(a)--(b) The intensity maps of the Josephson-ring polariton condensate in the scalar case changing in time, obtained from numerical GPE simulations for initial particle imbalance $z(0) = 0.280$ (a) and $z(0)=0.599$ (b). Top to bottom: at zero time, after half a period of oscillations, and after one oscillation period. In (a), $T=394$~ps and in (b), $T=176$~ps. In (b) the maximum of population never leaves the left half of the ring. (c) Wave functions of the localised states $\Phi_L = (\psi_{\rm g}+\psi_{\rm e})/\sqrt{2}$ and $\Phi_R = (\psi_{\rm g}-\psi_{\rm e})/\sqrt{2}$ formed from the superpositions of the ground and first excited states of the stationary GPE. (d)--(f) Evolution of the population imbalance $z(t)$ for the initial conditions as marked, obtained from the numerical simulations (the thin blue lines) and from the analytic two-mode model with the same initial conditions (the dashed black lines) and with corrected initial conditions (the thick yellow lines). The background gray lines show the relative phase $\delta(t)$ evolution in the respective regimes. Panels (d,~f) correspond to the intensity maps in (b,~c).
    Parameters: $m=10^{-5}m_0$, $\rho = 50~\mu$m$^{-2}$,  $g=0.2~\mu$eV~$\mu$m$^{2}$.
    (g) Dependence of the oscillations frequency on the initial particle imbalance, obtained from the two-mode model. The shaded area indicates the MQST regime. Blue: parameters as in (a--f), green: same but with $g=1~\mu$eV~$\mu$m$^{2}$. }
    \label{sfig1}
\end{figure*}

The superpositions of first two states of the stationary GPE $\psi_{\rm g}$ and $\psi_{\rm e}$  were used to form the desired initial conditions for our numerical simulations of the full nonlinear scalar GPE~\eqref{nGPE}.
The exemplary evolution of the condensate intensity $|\Psi({\bf r},t)|^2$ obtained from the simulations for different initial conditions is shown in Fig.~\ref{sfig1}(a,b), for the two distinctive regimes characteristic of the bosonic Josephson junction~\cite{gati2007bosonic}.
At any moment of time, from the GPE simulations one can define the populations of the left and right parts of the ring as $N_L(t) = \int_{0}^{\pi} |\psi(\phi,t)|^2 \,d\phi$ and $N_R(t) = \int_{\pi}^{2\pi} |\psi(\phi,t)|^2 \,d\phi$ and the integral particle imbalance
\begin{equation}
    z(t)=\frac{1}{N}[N_L(t)-N_R(t)],
\label{z(t)}
\end{equation}
where  the total number of particles $N =\pi \rho (r_2^2-r_1^2)$ is defined from the average condensate density $\rho$ on the ring.
The evolution of the population imbalance~\eqref{z(t)} obtained from the simulation of Eq.~\eqref{nGPE}  for three different values of initial imbalance $z(0)$ is shown in Fig.~\ref{sfig1}(d)---(f).

Apart from the population imbalance, in our simulations we look at the relative phase of the condensate wavefunction taken between the two half-rings. We track the phase of the order parameter at the middle of the respective areas:
\begin{equation}\label{delta}
    \delta(t) = \text{arg}[\psi(\pi/2,t)]-\text{arg}[\psi(3\pi/2,t)].
\end{equation}
Numerical evolution of the relative phase~\eqref{delta} for different values of $z(0)$ is shown in Fig.~\ref{sfig1}(d)---(f) on the gray-shaded background. At small $z(0)$ the oscillations of the population imbalance are harmonic, around the average value equal zero, while the relative phase oscillates also around zero with the same period and the amplitude not exceeding $\pi$. At the growth of the initial imbalance, the oscillations of populations become anharmonic and the relative phase oscillation amplitude tends to $\pi$. After a critical value of $z(0)$, the MQST regime is realized, which is when the populations imbalance oscillates around a new non-zero equilibrium position and the relative phase becomes running.

We note that GPE~\eqref{nGPE} provides yet another important insight. By varying the geometry of the barriers (their height $V$ and angular width $\alpha$) and looking at the resulting wavefunctions $\psi_{\rm g,e}$ and energy levels $E_{1,2,3}$ (similarly to Table~\ref{tab1}) at a fixed particle number, one can study their `quality' with respect to creating a BJJ.
As discussed previously, the natural lower limit for the barrier width arises from the comparison with the healing length at a given density; the upper boundary, on the other hand, is set by the required existence of the bound states within the wells. However, by modeling the GPE~\eqref{nGPE} we find that the restrictions on the barriers are more stringent. Fig.~\ref{fig:notun} shows three examples of geometries with ``allowed'' barriers width, which however do not present a system where BJJ is realized. In Fig.~\ref{fig:notun}(a), the barrier is too small to suppress the ground-state wavefunction and the condensate is not localized in either left or right half of the ring. In Fig.~\ref{fig:notun}(b) and (c), the barriers are either too wide or too high: in both of the considered cases, the wavefunctions are localized within the wells, but the tunneling is almost absent, resulting in basically constant particles populations $N_{L,R}$ on the two sides regardless of the initial conditions (see the insets). We conclude that for any chosen barriers strength $V$, the range of the barrier widths that are to be considered in our study is different. By performing this wavefunction analysis, we could identify the range of $\alpha$ for each $V$ in consideration, which is displayed symbolically in Fig.~\ref{fig5}.

\begin{figure}[t!]
    \centering
    \includegraphics[width=\linewidth]{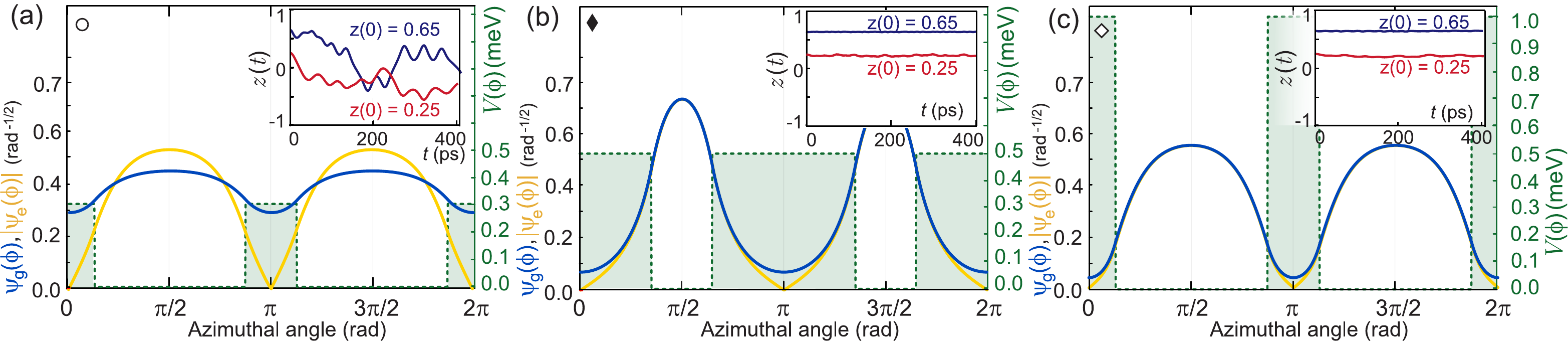}
    \caption{Examples of the ground-state (the blue lines) and first excited-state (the yellow lines) wavefunctions for barrier geometries {\it not} working as a Josephson junction. (a) $\alpha = 0.8$~rad, $V=0.3$~meV: the barrier is too weak to suppress the order parameter; (b) $\alpha = 2.2$~rad, $V=0.5$~meV: the barrier is too wide; (c) $\alpha = 0.8$~rad, $V=1.0$~meV: the barrier is too high. The insets show the corresponding behavior of the particle imbalance for $z(0)=0.25$ and 0.65. Symbols $\circ$, $\blacklozenge$ and $\diamond$ correspond to the same markers in Fig.~\ref{fig5}. For all panels, the one-component density $\rho=50~\mu$m$^{-2}$, $g=1~\mu$eV~$\mu$m$^{2}$.}
    \label{fig:notun}
\end{figure}

\section{Overview of the two-mode model}
\label{ssec2}

A standard approach to the theoretical description of bosonic Josephson phenomena in a double-well geometry is a two-mode model that describes the particle hopping between the two subsystems, regardless of their spatial distribution within the wells~\cite{raghavan1999coherent,milburn1997quantum}. We aim to apply the same approach to the polariton ring separated into two parts by the two junctions, and verify the applicability of the model comparing with the GPE numerical simulations described above. Importantly, we will build our two-mode model using the wave functions of the nonlinear problem (the GPE solutions) rather than linear Schr\"{o}dinger equation solutions.

We start our derivations from the second-quantized Hamiltonian in terms of creation and annihilation operators of particles in the left and right parts of the ring, $\hat{a}_L$ and $\hat{a}_R$. A similar approach for polariton double-well geometries was used in Ref.~\cite{shelykh2008josephson}. The 50:50 superpositions of the wavefunctions of the ground and first excited states obtained previously form (with a good accuracy) the states localized in the left $\Phi_L$ and right $\Phi_R$ parts of the ring: $\Phi_{L,R}(\phi) = (\psi_{\rm g}(\phi) \pm \psi_{\rm e}(\phi))/\sqrt{2}$ [see Fig.~\ref{sfig1}(c)]. For simplicity, we assume $\psi_{\rm g}(\phi)$, $\psi_{\rm e}(\phi)$ to be real. The field operator reads:
\begin{equation}
\hat{\Psi}(\phi) = \hat{a}_L \Phi_L(\phi) +  \hat{a}_R \Phi_R(\phi).
\end{equation}
The linear part of the Hamiltonian in second quantization is obtained in a standard way
\begin{equation}
\hat{\mathcal H}_0 = \int_{0}^{2\pi} \!\!\!\! d\phi \, \hat{\Psi}^{\dag}(\phi) \!\left[-\frac{\hbar^2}{2mR^2}\partial^2_{\phi\phi} + V(\phi)\right] \! \hat{\Psi}(\phi) = E_0\sum\limits_{i} \hat{a}_i^{\dag}\hat{a}_i - K_0 \sum\limits_{i} \hat{a}_{i}^{\dag}\hat{a}_{\overline{i}},
\end{equation}
where $i,j=\{L,R\}$, the bar denotes the opposite side of the ring, and we introduced the notation
\begin{equation}\label{E0K0}
        E_0, K_0 \equiv \frac12 \! \int_0^{2\pi} \!\!\!\! d\phi \left\{\psi_{\rm e}(\phi) \!\left[-\frac{\hbar^2}{2mR^2} \psi_{\rm e}^{\prime\prime}(\phi) + V(\phi)\psi_{\rm e}(\phi)\right] \pm  \psi_{\rm g}(\phi) \!\left[-\frac{\hbar^2}{2mR^2} \psi_{\rm g}^{\prime\prime}(\phi) + V(\phi)\psi_{\rm g}(\phi)\right] \! \right\}\!.
\end{equation}
In the non-interacting case ($g=0$), one would get $E_0 = (E_{\rm e} + E_{\rm g})/2$ and $K_0 = (E_{\rm e} - E_{\rm g})/2$.

The interaction part of the Hamiltonian is
\begin{equation}
    \hat{\mathcal H}_{int}=\frac{\tilde g}{2} \int_{0}^{2\pi} \!\!\!\! d\phi \, \hat{\Psi}^\dag(\phi) \hat{\Psi}^\dag(\phi) \hat{\Psi}(\phi) \hat{\Psi}(\phi).
    \label{Hint}
\end{equation}
Plugging the expressions for field operators into \eqref{Hint} and taking into account the bosonic commutation relations, after some algebra with even and odd wavefunctions $\psi_{\rm g}$ and $\psi_{\rm e}$ one gets:
\begin{multline}
    \hat{\mathcal H}_{int}= \frac{1}{4} \left[ (\kappa_{\rm gg} + 2\kappa_{\rm ge} + \kappa_{\rm ee}) \hat{N}^2 - 2(\kappa_{\rm gg} + 2\kappa_{\rm ge} + \kappa_{\rm ee}) \hat{N} + 4 \kappa_{\rm ge} (\hat{a}_R^{\dag} \hat{a}_R - \hat{a}_L^{\dag} \hat{a}_L)^2  \right.+\\
    \left. 2 (\kappa_{\rm gg}-\kappa_{\rm ee}) (\hat{N}-1) (\hat{a}_L^{\dag} \hat{a}_R + \hat{a}_R^{\dag} \hat{a}_L) + (\kappa_{\rm gg} - 2\kappa_{\rm ge} + \kappa_{\rm ee})(\hat{a}_L^{\dag} \hat{a}_R + \hat{a}_R^{\dag} \hat{a}_L)^2\right]\!,
\end{multline}
where $\hat{N}=\hat{a}_R^{\dag} \hat{a}_R + \hat{a}_L^{\dag} \hat{a}_L$ is the operator of full number of particles in the condensate, which is an integral of motion, and the coefficients are defined by the overlap integrals
$$\kappa_{\mu \nu} = \frac{\tilde g}{2}\int_{0}^{2\pi} \!\!\!\! d\phi \, \psi_\mu^2(\phi) \psi_\nu^2(\phi),\quad \mu,\nu = \{\rm g,e\}.$$

Adding the linear and nonlinear parts together and neglecting the terms of the order unity compared to the particle number, the Hamiltonian in the two-mode approximation reads
\begin{equation}
\hat{\mathcal H}=\frac{U}{2}(\hat{a}_R^{\dag} \hat{a}_R - \hat{a}_L^{\dag} \hat{a}_L)^2 - \left[K_0 + K_1 \hat{N}\right] \! (\hat{a}_L^{\dag} \hat{a}_R + \hat{a}_R^{\dag} \hat{a}_L) + \bigl[E_0 - \delta E_+\bigr] \hat{N} + \frac{\delta E_+}{2}\hat{N}^2 + \frac{\delta E_-}{2}(\hat{a}_L^{\dag} \hat{a}_R + \hat{a}_R^{\dag} \hat{a}_L)^2,
    \label{H2mod}
\end{equation}
where
\begin{equation}\label{notations}
U = \tilde g \int_{0}^{2\pi} \!\!\!\! d\phi \, \psi_{\rm g}^2 \psi_{\rm e}^2, \quad
 K_1 = \frac{\tilde g}{4} \int_{0}^{2\pi} \!\!\!\! d\phi \, (\psi_{\rm e}^4 - \psi_{\rm g}^4), \quad
\delta E_{\pm} =  \frac{\tilde g}{4} \int_{0}^{2\pi} \!\!\!\! d\phi \, (\psi_{\rm e}^4 \pm 2 \psi_{\rm e}^2 \psi_{\rm g}^2 + \psi_{\rm g}^4).
\end{equation}

Numerical evaluation shows that the last term in~\eqref{H2mod} is two orders of magnitude smaller than the other terms and can be neglected. In the Heisenberg picture, the particle operators depend on time and satisfy $i\hbar \partial_t \hat{a}_i = [\hat{a}_i, \hat{\mathcal H}]$ with the Hamiltonian~\eqref{H2mod}. Given the approximations above, we calculate the commutator neglecting the terms $\delta E_-$ and the terms of order of unity (compared to $N\gg 1$). Since the particle occupation of the two half-rings is macroscopic, it is justified to replace the particle operators with $C$--numbers  $\hat{a}_i \rightarrow \psi_i$, and obtain the set of coupled equations for the condensate order parameters $\psi_i$ of the left and right halves of the ring ($i=\{L,R\}$):
\begin{equation}\label{set_two}
\begin{split}
        i \hbar \partial_t\psi_L(t) =  \bigl(E_0 + \delta E_+ N\bigr)\psi_L + U (|\psi_L|^2-|\psi_R|^2)\psi_L - K_1 (|\psi_L|^2\psi_R + \psi_R^* \psi_L^2) - \Omega\psi_R, \\
        i \hbar \partial_t\psi_R(t) =  \bigl(E_0 + \delta E_+ N\bigr)\psi_R + U (|\psi_R|^2-|\psi_L|^2)\psi_R - K_1 (|\psi_R|^2\psi_L + \psi_L^* \psi_R^2) - \Omega\psi_L,
\end{split}
\end{equation}
where $\Omega = K_0 + K_1 N$.
The parameters governing the dynamics of Eqs.~\eqref{set_two} are listed in Table~\ref{tab2}.
\begin{table}[h!]
    \centering
    \begin{tabular}{c|c|c|c|c}
        \hline
        & $g = 0.2~\mu$eV~$\mu$m$^2$ &  $g = 1~\mu$eV~$\mu$m$^2$ &  $g = 1~\mu$eV~$\mu$m$^2$ &  $g = 1~\mu$eV~$\mu$m$^2$\\
        & $\rho = 50~\mu$m$^{-2}$ & $\rho = 25~\mu$m$^{-2}$ & $\rho = 50~\mu$m$^{-2}$ & $\rho = 100~\mu$m$^{-2}$  \\
        \hline\hline
        $N$ & 6911 & 3456 & 6911 & 13823 \\
        $E_0$ (meV) & 0.0379 & 0.0387 & 0.0409 & 0.0463 \\
        $K_0$ ($\mu$eV) & $1.14$ & $1.10$ & $0.957$ & $0.406$ \\
        \hline
        $U$ ($\mu$eV) & $3.429\times10^{-3}$ & $0.01641$ & $0.01554$ & $0.01444$ \\
        $\delta E_+$ ($\mu$eV) & $3.431\times10^{-3}$ & $0.01642$ & $0.01555$ & $0.01447$ \\
        $K_1$ ($\mu$eV) & $3.792\times10^{-5}$ & $2.071\times10^{-4}$ & $2.351\times10^{-4}$ & $2.885\times10^{-4}$\\
        \hline
        $E_0 + \delta E_+ N$ (meV) & 0.0616 & 0.0955 & 0.1484 & 0.2464 \\
        $\Omega$ ($\mu$eV) & 1.41 & 1.82 & 2.58 & 4.39 \\
        \hline\hline
        $\Lambda=UN/\Omega$ & 16.81 & 31.16 & 41.63 & 45.47 \\
        $z^{\rm crit} = (2/\Lambda)\sqrt{\Lambda-1}$ & 0.473 & 0.352 & 0.306 & 0.293 \\
        \hline
    \end{tabular}
    \caption{The two-mode model coefficients~\eqref{notations} for considered values of $g$ and the average one-component  density $\rho$.
    }
    \label{tab2}
\end{table}

Importantly, since we consider a fully-nonlinear problem when calculating the wave functions, the coefficients in the Eqs.~\eqref{set_two} are dependent both on the geometry of the barriers and on the density of the particles. To illustrate this, in Fig.~\ref{fig_density} we plot the parameters which are most sensitive to the density change as functions of the barrier width $\alpha$. One sees that while the overlap of the ground-state and excited-state wave functions [in $U$, see Fig.~\ref{fig_density}(a)] grows with $\alpha$, at the same time it gets slightly lower with the growth of the density. At the same time, the parameter $K_0$ rapidly drops with the growth of $\alpha$ [see Fig.~\ref{fig_density}(b)] and can even become negative for wide barriers and large densities, hence the tunneling  constant $\Omega = K_0 + K_1N$ starts to experience competition between the two terms  at the increase of the particle number [for the behaviour of $K_1N$ versus the barrier width, see Fig.~\ref{fig5}(d)].
\begin{figure}[t!]
    \centering
    \includegraphics[width=0.6\linewidth]{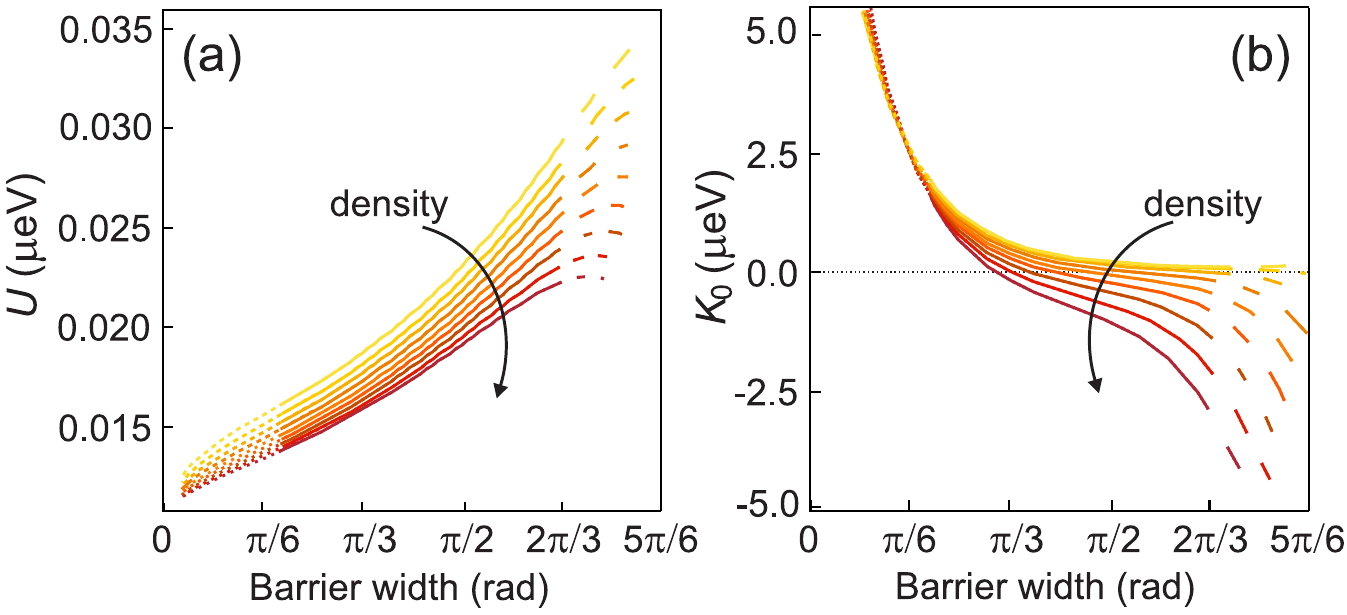}
    \caption{Dependence of the overlap integrals $U$~\eqref{notations} and $K_0$~\eqref{E0K0} on the barrier width $\alpha$ for one-component density varying from $10~\mu$m$^{-2}$ (yellow) to $90~\mu$m$^{-2}$ (dark red), at a fixed barrier height $V=0.5$~meV and interaction constant $g=1~\mu$eV~$\mu$m$^{2}$. The dotted and dashed part of the lines correspond to irrelevant barrier widths (see the discussion around Fig.~\ref{fig:notun}).}
    \label{fig_density}
\end{figure}

To compare the results obtained from Eqs.~\eqref{set_two} in terms of the particle imbalance evolution with the results obtained from full numerical spatio-temporal simulation of the GPE~\eqref{nGPE}, we substitute $\psi_i(t)=\sqrt{N_i(t)}e^{i\delta_i(t)}$ in Eqs.~\eqref{set_two} and separate the real and imaginary parts. The we get for the population imbalance $z=(N_L-N_R)/N$ and the phase difference of the opposite parts of the ring $\delta = \delta_R-\delta_L$ the equations
\begin{equation}
    \begin{split}
        &\dot{z} = -\frac{2\Omega}{\hbar}\sqrt{1-z^2} \sin{\delta} \\
        &\dot{\delta}= 2\, \frac{UN}{\hbar}z+\frac{2\Omega}{\hbar}\frac{z}{\sqrt{1-z^2}} \cos{\delta},
    \end{split}
    \label{two mod z delta}
\end{equation}
which up to the coefficients values and the energy detuning term coincides with the usual model for the double-well bosonic Josephson junction~\cite{raghavan1999coherent,gati2007bosonic,milburn1997quantum}. Eqs.~\eqref{two mod z delta} transparently define the characteristic time scale of the problem $\hbar/2\Omega$ and the interaction-to-tunneling ratio $\Lambda=UN/\Omega$.
Provided $\Lambda>1$, the fixed points of this Hamiltonian system $\dot{z}=-\partial\mathscr{H}/\partial\delta$, $\dot{\delta} = \partial\mathscr{H}/\partial z$ with $\mathscr{H} = (2\Omega/\hbar)[\Lambda z^2/2 - \sqrt{1-z^2}\cos\delta]$ are centers $\{z=0,\delta= 2\pi k\}$ and $\{z=\pm\sqrt{1-1/\Lambda^2},\delta=(2k+1)\pi\}$ ($k$ being integers) corresponding to the so-called zero- and $\pi$-phase modes (see~\cite{raghavan1999coherent}).
From the values listed in Table~\ref{tab2}, one can see that with the growth of the nonlinearity, the Josephson dynamics rapidly becomes faster: considering $\rho = 100~\mu$m$^{-2}$, for $g = 0.2~\mu$eV~$\mu$m$^2$ the timescale $\hbar/2\Omega \simeq 150$~ps, whereas for $g=1~\mu$eV~$\mu$m$^2$ and the same density it drops to 44~ps. Furthermore, at a fixed $\Lambda$ and fixed initial $\delta(0)$, when $z(0)$ exceeds a critical value $z^{\rm crit} = \pm(2/\Lambda)\sqrt{\Lambda-1}$, the populations become macroscopically self-trapped with $\langle z\rangle\neq 0$ and a running relative phase.
As $\Lambda$ grows, the system experiences the MQST regime at smaller initial particle imbalances.
It is important to note that since our model is fully nonlinear, the critical value $z^{\rm crit}$ becomes also dependent on the nonlinearity via the values of $U$ and $\Omega$ [see Table~\ref{tab2} and Fig.~\ref{sfig1}(g)].

Solutions for zero initial phase difference and various initial imbalances are shown in Fig.~\ref{sfig1}(d)--(f) together with the numerical solution of the full Gross-Pitaevskii equation~\eqref{nGPE}. As one can see,  in terms of the particle imbalance, the results of the two-mode model in which the coordinate dynamics is fully integrated over, coincides with the results of the GPE numerical simulation. A small discrepancy in the oscillation frequencies is corrected by tuning of the initial conditions within a few \%.

\vspace{-10pt}
\section{Generalization to the spinor case}
\label{ssec3}

We now generalize the same approach for the case when the pseudospins of the particles are considered. The field operator $\hat{\Psi}(\phi)$ in the spinor case is defined as a sum of four terms
\begin{equation}\label{Psi_spinor}
    \hat{\Psi}(\phi) = \! \sum\limits_{i}  \sum\limits_{\sigma} \Phi_{i\sigma}(\phi) \hat{a}_{i\sigma} =\!  \begin{pmatrix}
       \Phi_L(\phi) \\ 0
    \end{pmatrix} \! \hat{a}_{L \uparrow} + \! \begin{pmatrix}
        0 \\ \Phi_L(\phi)
    \end{pmatrix} \! \hat{a}_{L \downarrow} + \! \begin{pmatrix}
        \Phi_R(\phi) \\ 0
    \end{pmatrix} \! \hat{a}_{R \uparrow} +  \! \begin{pmatrix}
        0 \\ \Phi_R(\phi)
    \end{pmatrix} \! \hat{a}_{R \downarrow},
\end{equation}
where $i=\{L,R\}$ and $\sigma=\{\uparrow,\downarrow\}$, while $\Phi_{L}(\phi)$ and $\Phi_R(\phi)$ are the single-particle wavefunctions localized in the left and right parts of the ring (same as before, see Fig.~\ref{sfig1}(c)). In the absence of the TE-TM splitting and neglection of the opposite spins interaction, the behavior of the two components of the condensate is independent. Thus we can use the Hamiltonian~\eqref{H2mod} endowing each mode with a spin index:
\begin{equation}
  \hat{\mathcal H} \!=\!\! \sum\limits_{\sigma} \! \left[\frac{U_\sigma}{2}(\hat{a}_{R \sigma}^{\dag} \hat{a}_{R \sigma} \!\!-\! \hat{a}_{L \sigma}^{\dag} \hat{a}_{L \sigma}\!)^{2} \!\!-\! \left(\!K_{0}^{\sigma} \!+\! K_{1}^{\sigma} \hat{N}_\sigma\!\right)\!\!(\hat{a}_{L \sigma}^{\dag} \hat{a}_{R \sigma} \!+\! \hat{a}_{R \sigma}^{\dag} \hat{a}_{L \sigma}\!) \!+\! \bigl(\!E_{0}^{\sigma} \!-\! \delta E_{+}^{\sigma}\bigr) \!\hat{N}_\sigma \!+\! \frac{\delta E_{+}^{\sigma}}{2}\hat{N}_\sigma^2 \!+\! \frac{\delta E_{-}^{\sigma}}{2}(\hat{a}_{L\sigma}^{\dag} \hat{a}_{R\sigma} \!+\! \hat{a}_{R\sigma}^{\dag} \hat{a}_{L\sigma}\!)^{2}\right]
\end{equation}
with $\hat{N}_\sigma = \hat{a}_{R \sigma}^{\dag} \hat{a}_{R \sigma} + \hat{a}_{L \sigma}^{\dag} \hat{a}_{L\sigma}$ denoting the operator of the number of particles with a given pseudospin projection. The coefficients $E_{0}^{\sigma}$, $K_{0}^{\sigma}$, $U_\sigma$, $K_{1}^{\sigma}$ and $\delta E_{\pm}^{\sigma}$ are defined the same way as before~\eqref{notations}, but are endowed with the spin index since, in the general case, the two pseudospin components may have different densities.

Considering the off-diagonal part of the Hamiltonian~\eqref{hamiltonian_spinor}, we derive the operator responsible for the spin flips due to the TE-TM splitting:
$$    \hat{\mathcal H}_{\rm TE-TM} = \int_{0}^{2\pi} \!\!\!\! d\phi \, \hat{\Psi}^\dag(\phi) \!
    \begin{pmatrix}
        0 & \Delta e^{-2i\phi}  \\
        \Delta e^{2i\phi} &  0
    \end{pmatrix}
    \! \hat{\Psi}(\phi),
$$
which with substitution of Eq.~\eqref{Psi_spinor} yields:
\begin{equation}\label{J}
    \hat{\mathcal H}_{\rm TE-TM} = J_{\uparrow \downarrow} (\hat{a}_{L\uparrow}^\dag \hat{a}_{L\downarrow} + \hat{a}_{R\uparrow}^\dag \hat{a}_{R\downarrow}) + J_{\downarrow \uparrow} (a_{L\downarrow}^\dag \hat{a}_{L\uparrow} + \hat{a}_{R\downarrow}^\dag \hat{a}_{R\uparrow}), \quad J_{\uparrow \downarrow} = J_{\downarrow \uparrow}^* = \frac{\Delta}{2} \int_{0}^{2\pi} \!\!\!\! d\phi \left[|\psi_{\rm e}(\phi)|^2 + |\psi_{\rm g}(\phi)|^2\right] e^{2i\phi}.
\end{equation}
Therefore the full Hamiltonian of the two-mode model in the spinor case has the form:
\begin{multline}\label{hamiltonian_spinor_app}
    \hat{\mathcal H}= \sum\limits_{\sigma} \left[\frac{U_\sigma}{2}(\hat{a}_{R \sigma}^{\dag} \hat{a}_{R \sigma} - \hat{a}_{L \sigma}^{\dag} \hat{a}_{L \sigma})^2 - \left[K_{0}^{\sigma} + K_{1}^{\sigma} \hat{N}_\sigma\right](\hat{a}_{L \sigma}^{\dag} \hat{a}_{R \sigma} + \hat{a}_{R \sigma}^{\dag} \hat{a}_{L \sigma}) \right.\\
    \left. + \bigl[E_{0}^{\sigma} - \delta E_{+}^{\sigma}\bigr] \hat{N}_\sigma + \frac{\delta E_{+}^{\sigma}}{2}\hat{N}_\sigma^2 + \frac{\delta E_{-}^{\sigma}}{2}(\hat{a}_{L\sigma}^{\dag} \hat{a}_{R\sigma} + \hat{a}_{R\sigma}^{\dag} \hat{a}_{L\sigma})^2 + J_{\sigma \bar{\sigma}} (\hat{a}_{L\sigma}^\dag \hat{a}_{L\bar{\sigma}} + \hat{a}_{R \sigma}^\dag \hat{a}_{R\bar{\sigma}}) \right]\!.
\end{multline}
Analogously to the scalar case, writing the Heisenberg equations for the particle operators in the left and right parts of the ring and integrating out the spatial dependencies of the wavefunctions, we obtain the set of four equations for the wave functions of the spin-up and spin-down condensates in the left and right parts of the ring:
\begin{equation}\label{set_four}
    \begin{split}
        i \hbar \partial_t\psi_{L\uparrow}(t) = E_0^\uparrow\psi_{L\uparrow} + \Bigl[(U_\uparrow + \delta E_+^\uparrow) |\psi_{L\uparrow}|^2 - K_1^\uparrow \psi_{R\uparrow}^* \Bigr. & \Bigl. \psi_{L\uparrow} \Bigr] \psi_{L\uparrow} - \Bigl[K_0^\uparrow + 2K_1^\uparrow |\psi_{L\uparrow}|^2 + K_1^\uparrow|\psi_{R\uparrow}|^2\Bigr] \psi_{R\uparrow}\\
        &  + \delta E_-^\uparrow\bigl(2|\psi_{R\uparrow}|^2\psi_{L\uparrow} + \psi_{R\uparrow}^2 \psi_{L\uparrow}^*\bigr) + J_{\uparrow\downarrow} \psi_{L \downarrow}, \\[5pt]
        i \hbar \partial_t\psi_{L\downarrow}(t) = E_0^\downarrow\psi_{L\downarrow} + \Bigl[(U_\downarrow + \delta E_+^\downarrow) |\psi_{L\downarrow}|^2 - K_1^\downarrow \psi_{R\downarrow}^* \Bigr. & \Bigl. \psi_{L\downarrow} \Bigr] \psi_{L\downarrow} - \Bigl[K_0^\downarrow + 2K_1^\downarrow |\psi_{L\downarrow}|^2 + K_1^\downarrow|\psi_{R\downarrow}|^2\Bigr] \psi_{R\downarrow}\\
        &  + \delta E_-^\downarrow\bigl(2|\psi_{R\downarrow}|^2\psi_{L\downarrow} +  \psi_{R\downarrow}^2 \psi_{L\downarrow}^*\bigr) + J_{\downarrow\uparrow} \psi_{L \uparrow},\\[5pt]
        i \hbar \partial_t\psi_{R\uparrow}(t) = E_0^\uparrow\psi_{R\uparrow} + \Bigl[(U_\uparrow + \delta E_+^\uparrow) |\psi_{R\uparrow}|^2 - K_1^\uparrow \psi_{L\uparrow}^* \Bigr. & \Bigl. \psi_{R\uparrow} \Bigr] \psi_{R\uparrow} - \Bigl[K_0^\uparrow + 2K_1^\uparrow |\psi_{R\uparrow}|^2 + K_1^\uparrow|\psi_{L\uparrow}|^2\Bigr] \psi_{L\uparrow}\\
        &  + \delta E_-^\uparrow\bigl(2|\psi_{L\uparrow}|^2\psi_{R\uparrow} + \psi_{L\uparrow}^2 \psi_{R\uparrow}^*\bigr) + J_{\uparrow\downarrow} \psi_{R \downarrow},\\[5pt]
        i \hbar \partial_t\psi_{R\downarrow}(t) = E_0^\downarrow\psi_{R\downarrow} + \Bigl[(U_\downarrow + \delta E_+^\downarrow) |\psi_{R\downarrow}|^2 -  K_1^\downarrow \psi_{L\downarrow}^* \Bigr. & \Bigl. \psi_{L\downarrow} \Bigr] \psi_{R\downarrow} - \Bigl[K_0^\downarrow + 2K_1^\downarrow |\psi_{R\downarrow}|^2 + K_1^\downarrow|\psi_{L\downarrow}|^2\Bigr] \psi_{L\downarrow}\\
        &  + \delta E_-^\downarrow\bigl(2|\psi_{L\downarrow}|^2\psi_{R\downarrow} +  \psi_{L\downarrow}^2 \psi_{R\downarrow}^*\bigr) + J_{\downarrow\uparrow} \psi_{R \uparrow}.
    \end{split}
\end{equation}
Compared to Eqs.~\eqref{set_two}, here we have regrouped the terms in a slightly different fashion since the numbers of particles with a spin projection $\sigma$ on the whole ring $N_\sigma = |\psi_{L\sigma}|^2 + |\psi_{R\sigma}|^2$ are not constant any longer. As before, the terms proportional to $\delta E_{-}$ are negligible. Noticing that $U = \delta E_+ - \delta E_-$ [see \eqref{notations}] and dropping $\delta E_{-}$ (which implies $\delta E_+ \approx U$), one arrives at the set of equations given in Eq.~\eqref{seteq}:
\begin{equation}
    \begin{split}
        i \hbar \partial_t\psi_{L\uparrow}(t) & = E_0^\uparrow\psi_{L\uparrow} + \Bigl(2U_\uparrow|\psi_{L\uparrow}|^2 - K_1^\uparrow \psi_{R\uparrow}^*\psi_{L\uparrow} \Bigr)\psi_{L\uparrow} - \Bigl[K_0^\uparrow + 2K_1^\uparrow |\psi_{L\uparrow}|^2 + K_1^\uparrow|\psi_{R\uparrow}|^2\Bigr] \psi_{R\uparrow} + J_{\uparrow\downarrow} \psi_{L \downarrow}, \\[5pt]
        i \hbar \partial_t\psi_{L\downarrow}(t) & = E_0^\downarrow\psi_{L\downarrow} + \Bigl(2U_\downarrow|\psi_{L\downarrow}|^2 - K_1^\downarrow \psi_{R\downarrow}^*\psi_{L\downarrow} \Bigr)\psi_{L\downarrow} - \Bigl[K_0^\downarrow + 2K_1^\downarrow |\psi_{L\downarrow}|^2 + K_1^\downarrow|\psi_{R\downarrow}|^2\Bigr] \psi_{R\downarrow} + J_{\downarrow\uparrow} \psi_{L \uparrow},\\[5pt]
        i \hbar \partial_t\psi_{R\uparrow}(t) & = E_0^\uparrow\psi_{R\uparrow} + \Bigl(2U_\uparrow|\psi_{R\uparrow}|^2 - K_1^\uparrow \psi_{L\uparrow}^* \psi_{R\uparrow} \Bigr)\psi_{R\uparrow} - \Bigl[K_0^\uparrow + 2K_1^\uparrow |\psi_{R\uparrow}|^2 + K_1^\uparrow|\psi_{L\uparrow}|^2\Bigr] \psi_{L\uparrow} + J_{\uparrow\downarrow} \psi_{R \downarrow}, \\[5pt]
        i \hbar \partial_t\psi_{R\downarrow}(t) & = E_0^\downarrow\psi_{R\downarrow} + \Bigl(2U_\downarrow|\psi_{R\downarrow}|^2 - K_1^\downarrow \psi_{L\downarrow}^* \psi_{R\downarrow} \Bigr)\psi_{R\downarrow} - \Bigl[K_0^\downarrow + 2K_1^\downarrow |\psi_{R\downarrow}|^2 + K_1^\downarrow|\psi_{L\downarrow}|^2\Bigr] \psi_{L\downarrow} + J_{\downarrow\uparrow} \psi_{R \uparrow}.
    \end{split}
\end{equation}

We note that the Hamiltonian~\eqref{hamiltonian_spinor_app} and the Eqs.~\eqref{set_four} were derived rigorously and that they differ from the phenomenologically-formulated Hamiltonian and equations for the order parameter presented in Ref.~\cite{shelykh2008josephson}.
In particular, since $\rho_\uparrow\neq\rho_\downarrow$, the two pseudospin components now have the energy detuning, while all the coefficients determining the dynamics are generally not constant. The interaction-related nonlinearity (the round bracket in each equation) acquires a small correction because of the tunneling; more importantly, the tunneling rates between the left and right parts of the ring (the square brackets in each equation)  depend nonlinearly on the density of each pseudospin component of the condensate (cf. the scalar case where the tunneling rates were constant and defined by the total number of particles). It is important to note that while the parameters $U_\sigma$, $K_1^\sigma$ and $J_{\sigma\bar{\sigma}}$ vary just slightly for the considered range of densities (see the three rightmost columns in Table~\ref{tab2} for $g=1~\mu$eV~$\mu$m$^2$), the change of the values $E_0^\sigma$ and $K_0^\sigma$ which produce the energy detuning between the components cannot be neglected. The tunneling dynamics is now very much affected by the internal Josephson effect. The spin-flip rate $J_{\uparrow\downarrow}$ according to Eq.~\eqref{J} is provided in Table~\ref{tab3}.

\begin{table}[h!]
    \centering
    \begin{tabular}{c|c|c|c|c}
        \hline
        & $g = 0.2~\mu$eV~$\mu$m$^2$ &  $g = 1~\mu$eV~$\mu$m$^2$ &  $g = 1~\mu$eV~$\mu$m$^2$ &  $g = 1~\mu$eV~$\mu$m$^2$\\
        & $\rho_\uparrow+\rho_\downarrow = 100~\mu$m$^{-2}$ & $\rho_\uparrow+\rho_\downarrow = 50~\mu$m$^{-2}$ & $\rho_\uparrow+\rho_\downarrow = 100~\mu$m$^{-2}$ & $\rho_\uparrow+\rho_\downarrow = 200~\mu$m$^{-2}$  \\
        \hline\hline
        $N_{tot} = N_\uparrow + N_\downarrow$ & 13823 & 6911 & 13823 & 27646 \\
         $J_{\uparrow\downarrow}$ (meV) & $-0.01076$ & $-0.01011$ & $-0.00922$ & $-0.00790$ \\
         \hline
         $\Lambda_{\rm int}$ & 2.15 & 5.47 & 11.3 & 24.31 \\
         $\wp_c^{\rm crit}$ & 0.998 & 0.773 & 0.568 & 0.397  \\
         \hline
    \end{tabular}
    \caption{The the total number of particles in the two components and the spin-flip rate $J_{\uparrow\downarrow}$ according to Eq.~\eqref{J} with $\Delta = 0.02$~meV, for different values of $g$ and the average total (two-component) density $\rho$. The last two lines display the characteristic parameters of internal Josephson effect in absence of the tunneling dynamics (see text).
    }
    \label{tab3}
\end{table}

For the spinor case, there are several ways to parametrize the initial conditions. Since the total number of particles on the ring $N_{tot} = N_\uparrow + N_\downarrow = N_L + N_R = |\psi_{L\uparrow}|^2 + |\psi_{R\uparrow}|^2 + |\psi_{L\downarrow}|^2 + |\psi_{R\downarrow}|^2$ remains fixed, we will parametrize the initial conditions once again in terms of the initial particle imbalance between the two half-rings $z(0)$ (defined as in Eq.~\eqref{z(t)}, independent of pseudospin) and the degree of circular polarization (DCP) in each half-ring $\wp_c^{L,R}(0)$:
\begin{equation}
    \wp_c^L(t) = \frac{|\psi_{L\uparrow}(t)|^2 - |\psi_{L\downarrow}(t)|^2}{|\psi_{L\uparrow}(t)|^2 + |\psi_{L\downarrow}(t)|^2}, \quad \wp_c^R(t) = \frac{|\psi_{R\uparrow}(t)|^2 - |\psi_{R\downarrow}(t)|^2}{|\psi_{R\uparrow}(t)|^2 + |\psi_{R\downarrow}(t)|^2}.
\end{equation}
Then initial conditions are expressed as follows:
\begin{equation}
    \begin{split}
        &|\psi_{L\uparrow}(0)|^2 = \frac{1 + \wp_c^L(0)}{2}\frac{1+z(0)}{2} N_{tot}\\
        &|\psi_{L\downarrow}(0)|^2 = \frac{1-\wp_c^L(0)}{2}\frac{1+z(0)}{2} N_{tot}
    \end{split}
    \quad\quad
    \begin{split}
        &|\psi_{R\uparrow}(0)|^2 = \frac{1 + \wp_c^R(0)}{2}\frac{1-z(0)}{2} N_{tot} \\
        &|\psi_{R\downarrow}(0)|^2 = \frac{1-\wp_c^R(0)}{2}\frac{1-z(0)}{2} N_{tot}.
    \end{split}
\end{equation}
In all the simulations presented, we settle for the case $\wp_c^L(0)=\wp_c^R(0)\equiv\wp_c(0)$ which corresponds the same degree of polarization along the whole ring at the beginning of the evolution. Results of numerical simulations of Eqs.~\eqref{set_four} are presented in Figs.~\ref{fig2}--\ref{fig4},~\ref{fig6} above and in Figs.~\ref{sfig3} and \ref{sfig2} below. \\

\begin{figure}[t!]
    \centering
    \includegraphics[width=0.72\linewidth]{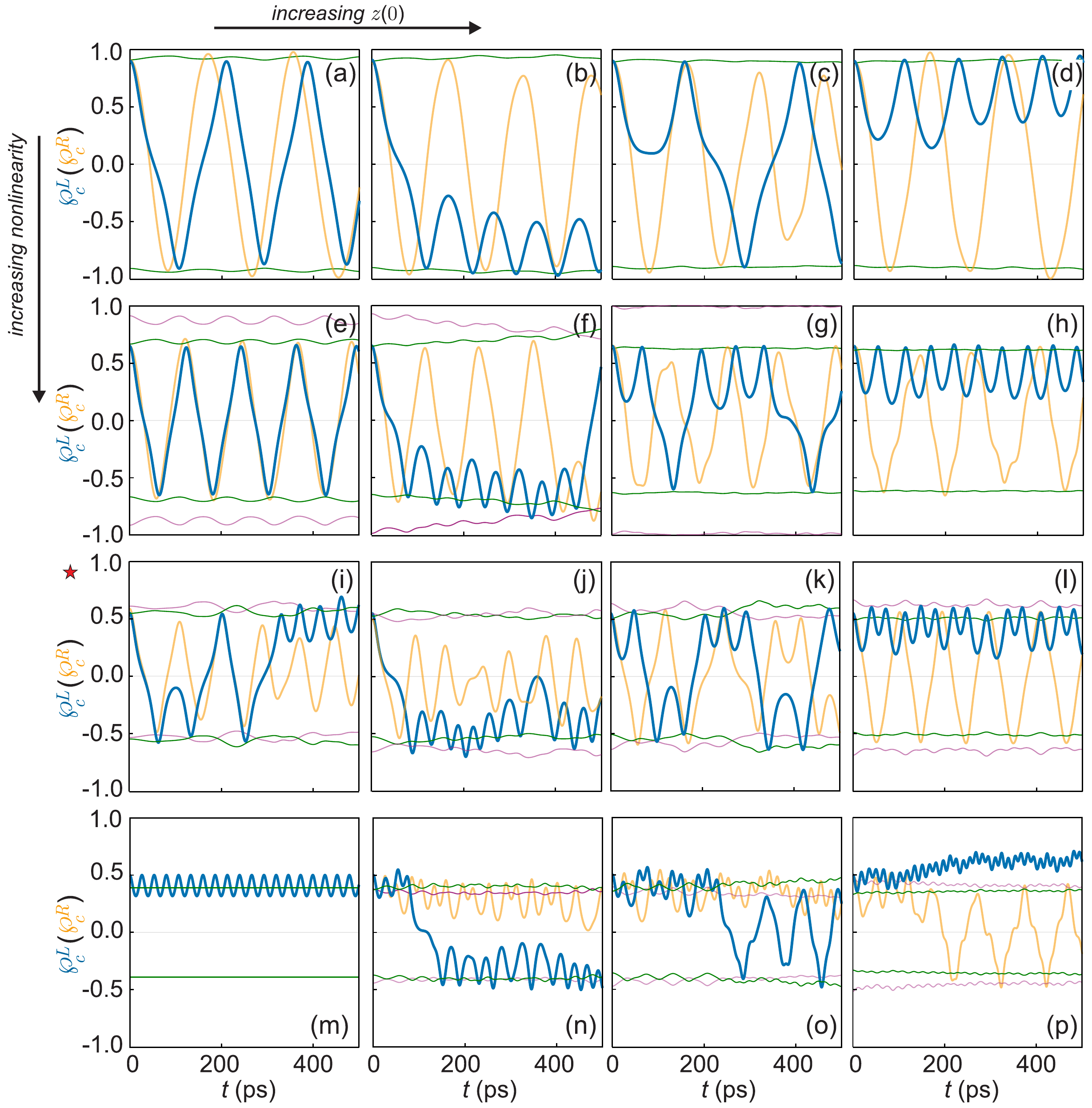}
    \caption{Oscillations of the DCP on the left (the blue lines) and right (the yellows lines) halves of the ring for different nonlinearities across the critical region. For all panels, $m = 10^{-5} m_0$ and $\Delta = 0.02$~meV. (a--d) $g = 0.2~\mu$eV~$\mu$m$^2$, $\rho_{tot} = 100~\mu$m$^{-2}$, $\wp_c(0)=0.9$
    (e--h) $g = 1~\mu$eV~$\mu$m$^2$, $\rho_{tot} = 50~\mu$m$^{-2}$, $\wp_c(0)=0.65$ (i--l) $g = 1~\mu$eV~$\mu$m$^2$, $\rho_{tot} = 100~\mu$m$^{-2}$, $\wp_c(0)=0.55$ (m--p) $g = 1~\mu$eV~$\mu$m$^2$, $\rho_{tot} = 200~\mu$m$^{-2}$, $\wp_c(0)=0.5$. The thin green (purple) lines show the critical values $\wp_c^{\rm crit}(t)$ evolution for the left (right) half-ring. Initial particle imbalance $z(0)$: (a) 0.55, (b) 0.6, (c) 0.65, (d) 0.7; (e) 0.4, (f) 0.5, (g) 0.6, (h) 0.7; (i) 0.05, (j) 0.15, (k) 0.25, (l) 0.35; (m) 0.0, (n) 0.1, (o) 0.2, (p) 0.4. See also the corresponding panels of Fig.~\ref{sfig2}.}
    \label{sfig3}
\end{figure}

\begin{figure}[t!]
\centering\includegraphics[width=0.75\textwidth]{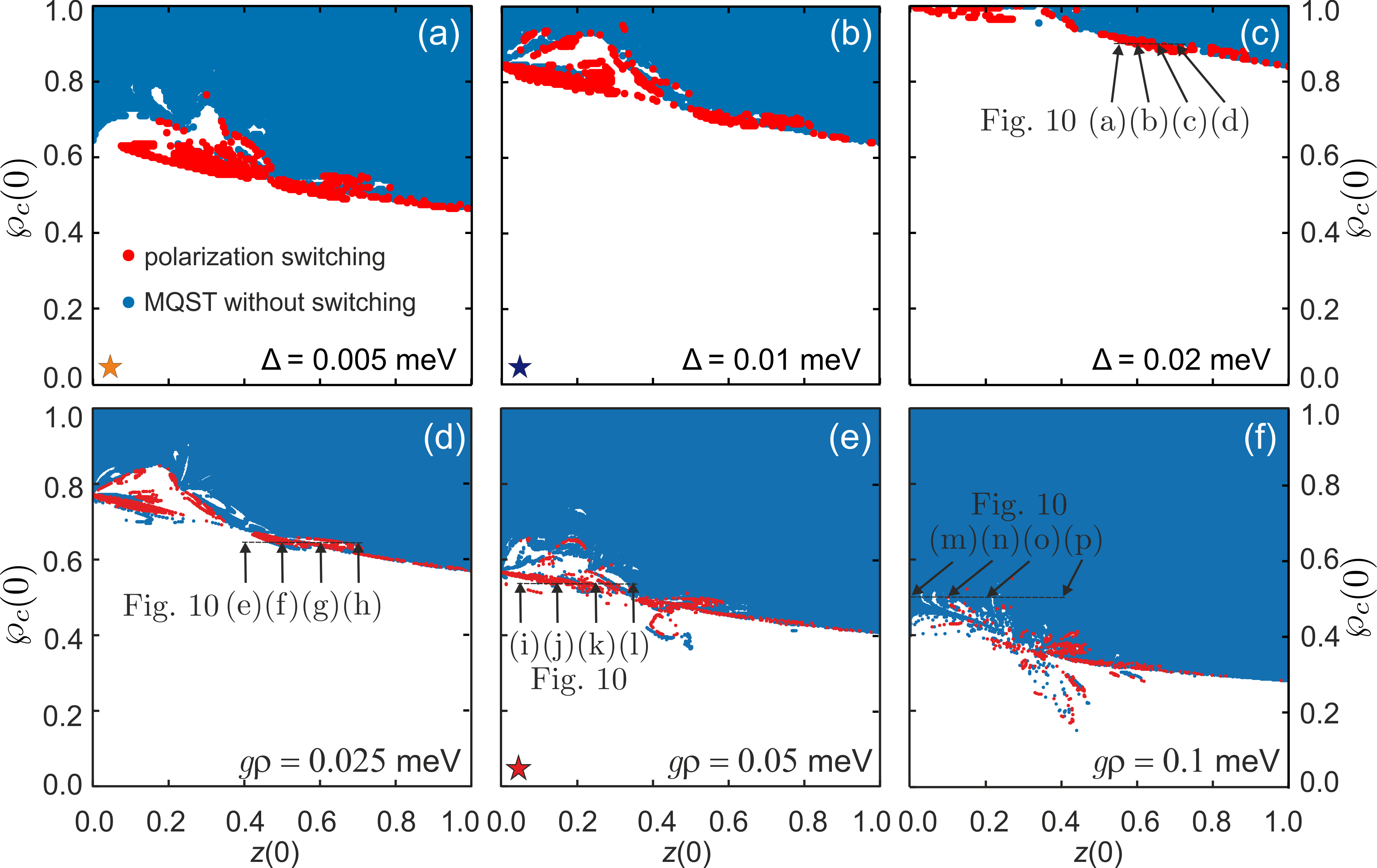}
\caption{\textbf{Oscillatory regimes diagrams depending on the TE-TM splitting $\Delta$ and the nonlinearity.} Regimes of $\wp_c^L(t)$ dynamics depending on the initial circular polarization degree $\wp_c(0)$ and population imbalance between the left and right half-rings $z(0)$: conventional oscillations (white regions), polarization switching (red points) and self-localization without switching (blue points).  Top row: $\rho_{tot} = 100~\mu$m$^{-2}$, $g=0.2~\mu$eV~$\mu$m$^2$ and the TE-TM splitting characterized by (a) $\Delta = 0.005$~meV, (b) $\Delta = 0.01$~meV, (c) $\Delta = 0.02$~meV. Bottom row: $\Delta = 0.02$~meV, $g=1~\mu$eV~$\mu$m$^2$ and (d)  $\rho_{tot} = 50~\mu$m$^{-2}$, (e)  $\rho_{tot} = 100~\mu$m$^{-2}$, (f)  $\rho_{tot} = 200~\mu$m$^{-2}$. The value of the average nonlinearity $g\rho$ marked on the panels is defined by half of the total density (since the polaritons of the opposite spins are assumed non-interacting). Panel (f) corresponds to Fig.~\ref{fig4}(b). In (c)--(f), small arrows indicate the parameters across the critical region at which the panels of Fig.~\ref{sfig3} are calculated.}
\label{sfig2}
\end{figure}

A simplification of equations~\eqref{int1}--\eqref{int3} for the internal Josephson effect (polarization dynamics) that allows to obtain the critical DCP value~\eqref{Pcrit} is the case of no spatial oscillations [i.e. $N_L(t) = N_R(t) = N_{tot}/2$ and $\delta_\uparrow (t) = \delta_\downarrow (t) = 0$]. Then one has $\wp_c^L = \wp_c^R = \wp_c$ subject to evolution equations
\begin{equation}\label{internal}
    \begin{split}
        \dot{\wp}_c & = \frac{2J}{\hbar} \sqrt{1-(\wp_c)^2} \sin\delta_{\downarrow\uparrow}, \\[5pt]
        \dot{\delta}_{\downarrow\uparrow}  & = 
        \frac{\mathscr{E}^{\uparrow\downarrow}}{\hbar} +
        \frac{(U - 2K_1)N_{tot}}{\hbar}\wp_c -\frac{2J}{\hbar}\frac{\wp_c}{\sqrt{1-(\wp_c)^2}}\cos\delta_{\downarrow\uparrow},
    \end{split}
\end{equation}
where the energy detuning between the two components is defined by the ground-state wavefunctions of the stationary GPE at different particle numbers:
\begin{equation}\label{dEg}
        \mathscr{E}^{\uparrow\downarrow} =  \int_0^{2\pi} \!\!\!  \psi_{\rm g}^\uparrow(\phi) \!\left[-\frac{\hbar^2}{2mR^2} \psi_{\rm g}^{\uparrow\,\prime\prime}(\phi) + V(\phi)\psi_{\rm g}^\uparrow(\phi)\right] \! d\phi  -  \int_0^{2\pi} \!\!\!  \psi_{\rm g}^\downarrow(\phi) \!\left[-\frac{\hbar^2}{2mR^2} \psi_{\rm g}^{\downarrow\,\prime\prime}(\phi) + V(\phi)\psi_{\rm g}^\downarrow(\phi)\right] \! d\phi.
\end{equation}
For simplicity, here we also assumed that for the chosen total density range the coefficients $U$ and $K_1$ are approximately independent on the components' fractional densities during the dynamics: $U_\uparrow \approx U_\downarrow = U$, $K_1^\uparrow \approx K_1^\downarrow = K_1$ (see Table~\ref{tab2}). Comparing to Eqs.~\eqref{two mod z delta}, we see that in this case the characteristic time scale of internal Josephson dynamics is equal $\hbar/2J$, while the nonlinearity parameter is defined as $\Lambda_{\rm int} = (U - 2K_1)N_{tot}/2J$ (here `int' stands for `internal'). Contrary to the conventional two-mode dynamics, the timescale of oscillations is weakly dependent on the particle number (see the values for $J_{\uparrow\downarrow}$ in Table~\ref{tab1}) and is of the order of $\hbar/2J \sim 35$~ps. The critical value of the initial DCP to reach the polarization self-localization regime (an analog of MQST), $\wp_c^{\rm crit} = \pm (2/\Lambda_{\rm int}) \sqrt{\Lambda_{\rm int} - 1}$, on the other hand, drops quite rapidly: for $g\rho=0.02$~meV the self-localization of polarization is reached only at $\wp_c(0)>0.998$, while for $g\rho=0.1$~meV it is reached much sooner, at $\wp_c(0)>0.56$ [see Table~\ref{tab3} and Fig.~\ref{fig4}(a)].

Fig.~\ref{sfig3} shows additional examples of the polarization oscillatory dynamics across the critical region (i.e. when $z(0)$ is varied while $\wp_c(0)$ is in the vicinity of $\wp_c^{\rm crit}$). Each row of panels in Fig.~\ref{sfig3} corresponds to a different nonlinearity in the system (top to bottom, $g\rho$ increasing from $0.01$ to $0.1$~meV) and is to be understood together with 
Fig.~\ref{sfig2}.

\section{Diagrams of the oscillatory regimes and polarization switching} \label{ssec4}

Finally, we investigate how the magnitude of the TE-TM splitting $\Delta$ [controlled by the ring width $a$, see Fig.~\ref{fig1}(a)] and the average nonlinearity $g\rho$ influence the appearance of the polarization switching regime. To illustrate our findings, we created diagrams in the parameter space of initial conditions $\wp_c(0)$ and $z(0)$ [similar to Fig.~\ref{fig4}(b)], displaying the areas of conventional Josephson oscillations (white) and the self-localization (blue), together with the narrow region of polarization switching (red). All panels show the regimes of internal Josephson dynamics for the left half of the ring. The top row of Fig.~\ref{sfig2} shows such diagrams for three values of $\Delta = 0.005, \; 0.01, \;$ and $0.02$~meV (the ring widths $a=3.5$, 2.65, and $2~\mu$m, respectively),  for the average densities $\rho_\uparrow=\rho_\downarrow = 50~\mu$m$^{-2}$ (total density $\rho_{tot} = 100~\mu$m$^{-2}$) and $g=0.2~\mu$eV~$\mu$m$^2$. Noteworthy, the localization region shifts upwards to larger $\wp_c(0)$ as $\Delta$ increases, as the critial value $\wp_c^{\rm crit}$ grows, see Fig.~\ref{fig4}(a). The polarization switching regime preserves (but narrows) up to $\Delta = 0.03$~meV and then disappears, as the critical value $\wp_c^{\rm crit}$ is not reached any longer.

The bottom row of the same Figure displays the gradual disappearance of the narrow region corresponding to the switching regime with the growth of the nonlinearity. In Fig.~\ref{sfig2}(d)--(e), at the fixed interaction constant $g = 1~\mu$eV~$\mu$m$^2$ and $\Delta = 0.02$~meV, we increase the total density of the fluid from $\rho_{tot} = 50~\mu$m$^{-2}$ to $200~\mu$m$^{-2}$ (left to right). The critical value $\wp_c^{\rm crit}$ at which the transition to the self-localized polarization oscillations occurs rapidly drops (see Table~\ref{tab3}), and the initial conditions leading to the switching (the red points on the diagram) shrink and become more concentrated around the value $z(0) \sim z^{\rm crit}$ (in the center of the diagram), compared to the cases of smaller nonlinearity where they occurred at any $z(0)$ along the critical line corresponding to the transition to the self-localization regime (the boundary between the white and blue regions).
\end{widetext}


\end{document}